\documentclass[journal]{IEEEtran}
\usepackage{cite}
\usepackage{amsfonts}
\usepackage{amssymb}
\usepackage{amsmath}
\usepackage{soul}
\usepackage{xcolor}
\usepackage{mathrsfs}
\usepackage{stfloats}
\usepackage{bm}
\usepackage{graphicx}
\usepackage{subfigure}
\usepackage{color}

\IEEEaftertitletext{\vspace{-1.5\baselineskip}}

\newtheorem{theorem}{Theorem}

\newtheorem{lemma}{Lemma}
\newtheorem{remark}{Remark}

\newcounter{mytempeqncnt}

\begin{document}
%
\title{Spectral Efficiency Analysis of Cell-free Distributed Massive MIMO Systems with Imperfect Covariance Matrix}
%
%
%

\author{~Feng~Ye,~\IEEEmembership{Student Member,~IEEE,}~Jiamin~Li,~\IEEEmembership{Member,~IEEE,}~Pengcheng~Zhu,~\IEEEmembership{Member,~IEEE,}
	Dongming~Wang,~\IEEEmembership{Member,~IEEE,} and~Xiaohu~You,~\IEEEmembership{Fellow,~IEEE}
\thanks{\emph{Corresponding author: Jiamin Li (e-mail: lijiamin@seu.edu.cn)}}
\thanks{This work was supported in part by the National Natural Science Foundation of China (NSFC) under Grant 61971127, 61871465, 61871122, by the National Key Research and Development Program under Grant 2020YFB1806600, and by Natural Science Foundation of Jiangsu Province under Grant BK20180011.}
\thanks{J. Li, D. Wang and X. You are with the National Mobile Communications Research Laboratory, Southeast University, Nanjing 210096, China, and also with Purple Mountain Laboratories, Nanjing 211111, China (e-mail: lijiamin, wangdm, xhyu@seu.edu.cn).}
\thanks{F. Ye and P. Zhu are with National Mobile Communications Research Laboratory, Southeast University, Nanjing 210096, China (e-mail: yefeng, p.zhu@seu.edu.cn).}

}

%



\maketitle

\begin{abstract}
In this paper, the impacts of imperfect channel covariance matrix on the spectral efficiency (SE) of cell-free distributed massive multiple-input multiple-output (MIMO) systems are analyzed. We propose to estimate the channel covariance matrix by alternately using the assigned pilots and their phase-shifted pilots in different coherent blocks, which improves the accuracy of channel estimation with imperfect covariance matrix and reduces pilot overhead. Under this scheme, the closed-form expressions of SE with maximum ratio combination (MRC) and zero-forcing (ZF) receivers are derived, which enables us to select key parameters for better system performance. Simulation results verify the effectiveness of the proposed channel estimation method and the accuracy of the derived closed-form expressions. When more coherent blocks are used to estimate the covariance matrix, we can get better system performance. Moreover, some insightful conclusions are arrived at from the SE comparisons between different receiving schemes (ZF and MRC) and different pilot allocation schemes (orthogonal pilot and pilot reuse).
\end{abstract}



%
\IEEEpeerreviewmaketitle

\section{Introduction}
%
%
%
%

\IEEEPARstart{D}{ue} to the increasing requirements for high system throughput, low latency, ultra-reliability and near-instant connection, more innovative technologies are needed to pave the way for the development of B5G and 6G technologies\cite{shafi20175g}.
A large number of access points (APs) distributed in an area serve all users in the same time-frequency resource\cite{ngo2017cell,interdonato2019ubiquitous}.
Cell-free distributed massive MIMO can get all the benefits from massive MIMO (which facilitates propagation and channel strengthening when multiple antennas are used at the APs\cite{chen2017can}) and network MIMO (which increases macro-diversity gain). Therefore, it has a very high spectral efficiency, energy efficiency and coverage.
In addition, the cell-free structure can solve the problem that user performance and user location are related in the cellular network. Compared with the centralized system, the cell-free distributed massive MIMO system has advantages such as channel diversity, no switching, higher coverage and no need deploying cells in specific areas\cite{buzzi2017cell}.
Therefore, in recent years, cell-free distributed massive MIMO has aroused widespread research interest\cite{ngo2017total,zhang2017spectral,bashar2018cell,nguyen2017energy,hoang2018cell}.

Obtaining channel state information (CSI) is essential for taking advantage of the cell-free distributed massive MIMO system. There have been some studies on channel estimation methods.
In most of these studies, the minimum mean square error (MMSE) channel estimation method is extended to the cell-free distributed MIMO system, or make certain improvements on this basis \cite{mai2018cell,ngo2017total,dao2020effective,jin2019channel,wang2020uplink}.
There may be other channel estimation methods, such as least-squares (LS) channel estimation, but this results in very poor system performance\cite{mishra2015analysis}.
However, these estimations are performed under the premise that the covariance matrix is perfect. Actually, the covariance matrix is usually imperfect, and the imperfect covariance matrix will have an impact on system performance. Because covariance matrix information is important for resource allocation and to suppress pilot contamination, imperfect covariance matrix will cause serious pilot contamination and affect the system performance\cite{bjornson2016massive}.

Generally speaking,
there are six typical methods to obtain the covariance matrix:
1) Use a specific phase for learning the covariance matrix where every user in the network uses a unique orthogonal pilot, from which a sample correlation matrix can be formed, possibly by using regularization to get robustness\cite{yin2013coordinated};
2) Eliminate the interference by sending pilots from other interfering users to obtain the covariance matrix of the target user\cite{bjornson2016massive};
3) Add additional pilot sequences for covariance estimation in the coherent blocks to obtain accurate channel estimation\cite{upadhya2018covariance};
4) Instead of reserving a specific phase for learning the covariance matrix, the pilot assignment is changed between different coherence blocks\cite{neumann2018covariance};
5) Use the MMSE estimator of the structure model as the blueprint of the neural network structure, and a similarly effective (but suboptimal) estimator was obtained\cite{neumann2018learning};
6) Phase-rotate the pilot sequence in each coherence block to approximately decorrelate the channel observations at the APs\cite{interdonato2020self}.
However, these methods are all covariance matrix acquisition methods in a centralized cellular scenario, and there is no relevant research in a cell-free distributed massive MIMO scenario.

To meet the needs of super-dense user scenarios, the total number of antennas in cell-free massive MIMO systems need to be increased greatly.
With the increase of the total number of antennas in the system, the spatial multiplexing gain of the system will be greatly improved, and the number of users who can serve in the same frequency will also be greatly increased. However, the increase of system capacity is established when the CSI is known. If the orthogonal pilot is used to estimate the CSI, the pilot overhead will increase linearly with the increasing of the number of antennas.
If a completely orthogonal pilot is used, the pilot overhead will be very large, which ultimately restricts the improvement of the overall transmission efficiency.
For uplink wireless transmission, although the pilot overhead is only proportional to the number of users and independent of the number of antennas, it is still too overhead if the completely orthogonal pilot is adopted considering the large number of users the cell-free distributed massive MIMO needs to support.
Therefore, it is necessary to adopt a pilot multiplexing scheme to improve the utilization of pilot resources, which will result in pilot contamination\cite{haghighatshoar2017massive,marzetta2010noncooperative,jose2011pilot,pitarokoilis2017effect}.
Pilot contamination makes the estimation of the covariance matrix more complicated, because the users using the same pilot sequences will cause interference, so the channel estimates from which the covariance matrix estimates are obtained are themselves contaminated. Naively utilizing the contaminated channel estimates in a sample covariance estimator will result in the target user covariance matrix estimate containing the covariance matrices of the interference users, which in turn leads to inaccurate estimation of the covariance matrix and ultimately affects the overall transmission efficiency.

In this paper, we propose a new covariance matrix estimation method and derive the closed-form expressions for the uplink achievable rates in a cell-free distributed massive MIMO system with both MRC and ZF receiver under pilot contamination. The main contributions of this paper are summarized as follows:

\begin{enumerate}
 \setlength{\itemsep}{0pt}
 \setlength{\parsep}{0pt}
 \setlength{\parskip}{0pt}
\item We propose to estimate the channel covariance matrix by using alternate assigned pilots and their phase-shifted pilots in different coherent blocks, which obtains accurate channel estimation with low complexity and without extra pilots.
\item Given the estimated channel covariance matrix, we derived accurate closed-form expressions of the uplink achievable rates with both MRC and ZF receivers which enables the spectral efficiency analysis of cell-free distributed massive MIMO systems with imperfect covariance matrix.
\item Numerical simulations are performed to corroborate our theoretical analysis, and insightful conclusions are drawn from the comparison between system performance with perfect and imperfect covariance matrix using different receivers with different pilot allocation and the analyze of the relationship between the number of coherent blocks used to calculate the covariance estimation and the SE.
\end{enumerate}

The remainder of the paper is organized as follows. In Section II, we describe the system model including the system configuration, channel model, and channel estimation with pilot contamination in the case of perfect or imperfect covariance matrix. Section III contains pilot structure design and covariance matrix estimation.  The spectral efficiency is analyzed in section IV. Representative numerical results are given in Section V before we conclude the paper in Section VI.

{\emph{Notation}}: Boldface letters stand for matrices (upper case) or vectors (lower case). The transpose and conjugate transpose are denoted by ${\left(  \cdot  \right)^ \textrm{T} }$ and ${\left(  \cdot  \right)^ \textrm{H} }$ respectively. ${{\bf{I}}_{MN}}$ stands for the $M\times N$ identity
matrix, and $\mathcal{\mathcal{CN}}(\mu,\sigma^2)$ denotes the
circularly symmetric complex Gaussian distribution with mean $\mu$
and variance $\sigma^2$, while ${{\mathcal{W}}}(N,\mathbf{R})$ denotes Wishart matrix with $N$ degrees of freedom and $\mathbf{R}$ is the covariance matrix that corresponds to underlying Gaussian random vectors.

\section{System Model}
This paper considers a cell-free distributed massive MIMO system. There are $M$ APs equipped with $N$
antennas and $K$ single-antenna users.
Assuming that the system adopts a time division duplex (TDD) mode, the BS uses the reciprocity of the uplink and downlink channels to perform downlink multi-user precoding based on the uplink channel estimation results.

\subsection{Channel Model}
Considering a block fading channel, and the channel vector from user $k$ to all APs can be modeled as\cite{cao2018uplink}
\begin{equation}\label{ch_model}
{{\mathbf{g}}_k}{\rm{ = }}{\bf{\Lambda }}_k^{1/2}{{\mathbf{h}}_k}\in {{\mathbb{C}}^{MN\times {1}  }},
\end{equation}
with
\begin{equation}
\label{equ_channelM}
{{\bf{\Lambda }}_{{k}}} = {\rm{diag}}({[{\lambda _{k,1}} \cdots {\lambda _{k,M}}]^{\rm{T}}}) \otimes {{\mathbf{I}}_N}\in {{\mathbb{C}}^{MN\times {MN}  }},
\end{equation}
where ${\lambda _{km}}\buildrel \Delta \over = d_{k,m}^{ - \zeta }{s_{k,m}}$ represents the large-scale fading between user $k$ and AP $m$. ${d_{k,m}}$ is the distance from user $k$ to AP $m$, $\zeta $ is the path loss exponent, ${s_{k,m}}$ is a log-normal shadow fading variable. In addition, ${{\mathbf{h}}_k}\in {{\mathbb{C}}^{MN\times {1}  }}$ models small-scale fast fading, and each elements follows ${{\cal{CN}}}\left( {0,1} \right)$. So the channel ${{\mathbf{g}}_{k}}$ satisfies ${{\cal{N}}_{\rm{c}}}\left( {0,{\bm{\Lambda}}_k} \right)$, that is ${\mathbb{E}\left[ {{\mathbf{g}}_{k}} \right]=0}$, $\operatorname{cov}\left( {{\mathbf{g}}_{k}},{{\mathbf{g}}_{k}} \right)={{\mathbf{\Lambda}}_{k}}$.

The uplink received signal model is:
\begin{equation}
\label{equ_channel_r}
{\mathbf{y}=\sqrt{\rho }\sum\limits_{k=1}^{K}{{{\mathbf{g}}_{k}}{{\mathbf{x}}_{k}}}+\mathbf{n}},
\end{equation}
where ${\mathbf{y}}$ represents information received from all users, $\rho$ is the sending power, ${{\mathbf{x}}_{k}}$ is the sending information, $\mathbf{n}\sim {{\cal{N}}_{\rm{c}}}\left( 0,{{\sigma }^{2}} \right)$ is the noise and ${{\sigma }^{2}}$ is the noise power. We assume that ${{\mathbf{g}}_{k}}$ and $\mathbf{n}$ are independent of each other.

\subsection{Uplink Channel Estimation}

As widely recognized in most of articles, we first assume that the large-scale fading ${\bm{\Lambda }}_{k}$ is known to the BS, and the small-scale fading ${{\mathbf{h}}_{k}}$ needs to be estimated at the BS. Then, based on the actual situation that the covariance matrix is unknown, so the large-scale fading ${\bm{\Lambda }}_{k}$ is also unknown to the BS, and both the large-scale fading ${\bm{\Lambda }}_{k}$ and the small-scale fading ${{\mathbf{h}}_{k}}$ need to be estimated at the BS.

{\emph{1) Case I: Perfect covariance matrix} }

It is assumed that there are $P$ orthogonal pilot sequences of length ${\tau}$ to be used for uplink channel estimation, which is a constant independent of $K$, and the pilot signal is allocated to $K$ users when accessing the network. When $P<K$, there will be multiple users using the same pilot frequency, resulting in pilot contamination, which leads to the deterioration of the accuracy of channel estimation.
Let ${{\cal{U}}_{p}}\subset \left\{ 1,\ldots ,K \right\}$ denote the subset of users using the pilot $p$, all users using the $p$th pilot transmit the pilot signal ${{\mathbf{X}}_{p}}$ on the same time-frequency resource, and satisfy ${{\mathbf{X}}_{p}}\mathbf{X}_{p}^{\text{H}}={{\mathbf{I}}_{P}}$.
After correlating the received training signal with the conjugate of the pilot sequence, the BS estimates the channel based on the following observations
\begin{equation}
\label{equ_channel_m_e}
{{\mathbf{y}}_{p}}=\sqrt{\rho }\sum\limits_{i\in {{\cal{U}}_{p}}}{{{\mathbf{g}}_{i}}}+{{\mathbf{n}}_{p}}.
\end{equation}

Since ${{\mathbf{g}}_{k}}$ satisfies ${{\cal{N}}_{\rm{c}}}\left( {0,{\bm{\Lambda}}_k} \right)$, for $k\in{{\cal{U}}_{p}}$, the minimum mean square error (MMSE) estimate of channel ${{\mathbf{g}}_{k}}$ should be
\begin{align}
{{{\mathbf{\hat{g}}}}_{k}}=&\mathbb{E}\left[ {{\mathbf{g}}_{k}} \right]+\operatorname{cov}\left( {{\mathbf{g}}_{k}},{{\mathbf{y}}_{p}} \right)\operatorname{cov}{{({{\mathbf{y}}_{p}},{{\mathbf{y}}_{p}})}^{-1}}({{\mathbf{y}}_{p}}-\mathbb{E}\left[ {{\mathbf{y}}_{p}} \right])\nonumber\\
=&\frac{{\mathbf{\Lambda}_{k}}}{\sqrt{\rho }}{{\left( \sum\limits_{i\in {{\mathcal{U}}_{p}}}{{\mathbf{\Lambda}_{i}}}+\frac{{{\sigma }^{2}}}{\rho }{{\mathbf{I}}_{MN}} \right)}^{-1}}{{\mathbf{y}}_{p}},
\end{align}
with
\begin{equation}
\label{sample_cov_th}
{{{{\mathbf{\Sigma} }_{p}}=\operatorname{cov}\left( {{\mathbf{y}}_{p}},{{\mathbf{y}}_{p}} \right)=\rho \left( \sum\limits_{i\in {{\mathcal{U}}_{p}}}{{\mathbf{\Lambda}_{i}}}+\frac{{{\sigma }^{2}}}{\rho }{{\mathbf{I}}_{MN}} \right)}},
\end{equation}
where ${{\mathbf{\Sigma}}_{p}}$ is the covariance matrix of the received signal.

Therefore, for $k\in{{\cal{U}}_{p}}$, the channel estimation expression with perfect covariance matrix is
\begin{equation}
{{{\mathbf{\hat{g}}}_{k}}=\sqrt{\rho }{\mathbf{\Lambda }_{k}}{\mathbf{\Sigma}} _{p}^{-1}{{\mathbf{y}}_{p}}}.
\end{equation}

\emph{2) Case II: Imperfect covariance matrix}

In practice, the covariance matrix is usually imperfect, and the aforementioned channel estimation under the perfect covariance matrix does not meet actual requirements. At this point, both ${\mathbf{\Lambda }_{k}}$ and ${\mathbf{\Sigma}_{p}}$ are unknown quantities, so the channel estimation expression is transformed into
\begin{equation}\label{channel_es}
{{{\hat{\mathbf{g}}}_{k}}={\hat{\mathbf{\Lambda }}_{k}}{\hat{\mathbf{\Sigma}}} _{p}^{-1}{{\mathbf{y}}_{p}}}.
\end{equation}

Hence, if we want to obtain the accurate channel estimation with imperfect covariance matrix ,we must accurately estimate the covariance matrix. At the same time, it can also be seen in both (\ref{sample_cov_th}) and (\ref{channel_es}) that the covariance matrix cannot be calculated directly by simply using the contaminated channel, and the covariance matrix needs to be estimated separately.

Assuming that the channel is a block fading model, the channel remains unchanged within a certain coherence bandwidth ${{B}_{c}}$ and a certain coherence time ${{T}_{c}}$, that is, the channel remains unchanged within ${{\tau }_{c}}={{B}_{c}}{{T}_{c}}$ symbols. The covariance matrix is constant in the transmission bandwidth. Compared with the rapid change of the small-scale fading, it changes slowly in time, so it can be reasonably assumed that it remains unchanged within ${{\tau }_{s}}={{B}_{s}}{{T}_{s}}/{{B}_{c}}{{T}_{c}}={{B}_{s}}{{T}_{s}}/{{\tau }_{c}}$ coherent blocks. Therefore, channel estimation is required for each coherent block, but the covariance matrix only needs to be estimated once in ${{\tau }_{s}}$ coherent block. Therefore, according to this property, we estimate the covariance matrix and channel by using a special pilot structure.

In the following section, we propose a method of using different pilot in adjacent coherent blocks. Compared with the common method of adding extra pilots for covariance matrix estimation, this scheme can greatly reduce the pilot overhead without losing the estimation accuracy.

\section{Proposed Pilot Structure and Method for Estimating Covariance Matrix}

In this section, we first describe the pilot structure design. Then based on the structure, we estimate the sample covariance and individual covariance respectively.

\subsection{Pilot Structure Design}
Considering the block fading model, multiple coherent blocks can be used to estimate the covariance matrix. Since each coherent block sends pilots for channel estimation, in order not to add additional pilot, we consider using the channel estimated pilots for simultaneous covariance matrix estimation. However, due to pilot contamination, in order to obtain interference-free observations for each user, we should change the pilot structure. This pilot structure needs to be able to eliminate interference for accurate covariance matrix estimation without affecting channel estimation, so we can change the pilot by introducing random phase shift.

\begin{figure}
	\centering
	\includegraphics[width=8cm,height=5cm]{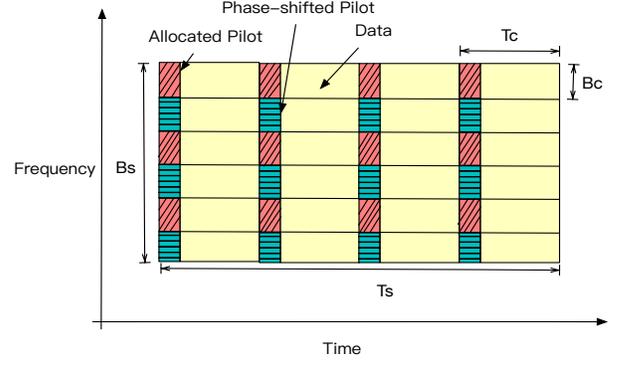}\\
	\caption{The structure of pilot in some coherent blocks with invariant covariance matrix}\label{fig_structure}
\end{figure}

In addition, in order to have a low complexity system, we should use a relatively simple pilot structure. So we consider a use of alternate pilots to estimate the covariance matrix, that is, each user alternately sends allocated pilot ${{\mathbf{X}}_{p}}$ and phase-shifted pilot ${\mathbf{\Phi }_{k}}$ in adjacent coherent blocks. ${\mathbf{\Phi }_{k}}$ is the pilot after the pilot ${{\mathbf{X}}_{p}}$ undergoes a random phase shift
\begin{equation}
{{\mathbf{\Phi}_{k}}\left[ 2n \right]={{e}^{j{{\theta }_{k,2n}}}}{{\mathbf{X}}_{p}}\left[ 2n-1 \right],n=1,2,\cdots},
\end{equation}
where $n$ is the index of coherent blocks, ${{\theta }_{k,2n}}$ is a random phase shift, which is irrelevant to the channel vector and noise and satisfies $\mathbb{E}\left[ {{e}^{j{{\theta }_{k,2n}}}} \right]=0$. So this random phase-shiftedt pilot ${\mathbf{\Phi }_{k}}$ is irrelevant to the channel and noise. The detailed structure of the pilot is depicted in Fig. \ref{fig_structure}.

Assuming that different users using the same pilot has different random phase shift ${{\theta }_{k,2n}}$. By observing the received signals of the alternate pilot, using the phase shift term irrelevance to remove interference from other users and noise, an accurate covariance matrix estimation can be obtained. In practical applications, ${{\theta }_{k,2n}}$ can be obtained by a pseudo-random sequence generator, and it is assumed that these phase shifts are known at the central processing unit (CPU).

\subsection{Sample Covariance Matrix Estimation}
Based on the pilot structure, the sample correlation matrix is used to approximate ${\mathbf{\Sigma}_{p}}$. We select the coherent blocks that send pilot ${{\mathbf{X}}_{p}}$ , that is, the interval coherent blocks are used to estimate ${\mathbf{\Sigma}_{p}}$. Assuming that the pilots of ${{N}_\mathbf{\Sigma }}\le {{\tau }_{s}}/2$ coherent blocks are received, the sample covariance matrix can be obtained as
\begin{equation}\label{sigma_hat}
{{{\hat{\mathbf{\Sigma}}}_{p}}=\frac{1}{{{N}_\mathbf{\Sigma }}}\sum\limits_{n=1}^{{{N}_\mathbf{\Sigma }}}{{{\mathbf{y}}_{p}}[2n-1]{{({{\mathbf{y}}_{p}}[2n-1])}^{\text{H}}}}},
\end{equation}
where ${{\mathbf{y}}_{p}}[2n-1]$ represents the received signal of the $(2n-1)$th coherent block. When ${{N}_\mathbf{\Sigma }}$ is larger, the estimated ${{\hat{\mathbf{\Sigma }}}_{p}}$ will be more accurate.

In fact, it is not necessary to select coherent blocks sending pilot ${{\mathbf{X}}_{p}}$, as long as the interval coherence blocks are selected to estimate the sample correlation matrix, the same result will be obtained.

\subsection{Individual Covariance Matrix Estimation}
In order to estimate individual covariance matrix ${{\hat{\mathbf{\Lambda} }}_{k}}$, adjacent coherent blocks are needed to remove the correlation. Based on the pilot sequence ${\mathbf{\Phi }_{k}}$, the pilot signal received by the APs from the users is:
\begin{equation}
{\mathbf{y}_{p}^{(\mathbf{\Phi} )}=\sqrt{\rho }\sum\limits_{i\in {{\mathcal{U}}_{p}}}{{{\mathbf{g}}_{i}}{{e}^{j{{\theta }_{i,2n}}}}}+{{\mathbf{n}}_{p}}}.
\end{equation}

According to the received signals of the alternate pilots, for $k\in {{\mathcal{U}}_{p}}$, the observed values of the received pilot signals of the two adjacent coherent blocks (i.e. coherent block pairs) are respectively
\begin{equation}
{\mathbf{\hat{h}}_{k}^{(1)}[2n-1]={{\mathbf{y}}_{p}}=\sqrt{\rho }\sum\limits_{i\in {{\mathcal{U}}_{p}}}{{{\mathbf{g}}_{i}}}+{{\mathbf{n}}_{p}}},
\end{equation}
and
\begin{align}
\mathbf{\hat{h}}_{k}^{(2)}[2n]=&{{\mathbf{y}}_{p}^{(\mathbf{\Phi})}}/{{e}^{j{{\theta }_{i,2n}}}}\nonumber\\
=&\sqrt{\rho }{\mathbf{g}}_{k}+\sqrt{\rho }\sum\limits_{{i}\ne{k},i\in {{\mathcal{U}}_{p}}}{{{\mathbf{g}}_{i}}}{{e}^{j({{\theta }_{i,2n}}-{\theta }_{k,2n})}}+{{\mathbf{n}}_{p}}{{e}^{j{{\theta }_{k,2n}}}},
\end{align}
where $\mathbf{\hat{h}}_{k}^{(1)}[2n-1]$ is obtained by (\ref{equ_channel_m_e}).

For $k\in {{\mathcal{U}}_{p}}$, the covariance matrix of the channel estimated according to the observations of the adjacent alternate pilot signals is
\begin{align} \operatorname{cov}(\mathbf{\hat{h}}_{k}^{(1)}[2n-1],\mathbf{\hat{h}}_{k}^{(2)}[n])=&\mathbb{E}\left[ \mathbf{\hat{h}}_{k}^{(1)}[2n-1]{{\left( \mathbf{\hat{h}}_{k}^{(2)}[n] \right)}^\text{H}} \right] \nonumber\\
=&\mathbb{E}\left[ \rho {{\mathbf{g}}_{k}}\mathbf{g}_{k}^{\text{H}} \right]\nonumber\\
=&\rho {{\mathbf{\Lambda}}_{k}}.
\end{align}
The equation is established because it is assumed that ${{\mathbf{g}}_{k}}$ and ${{\mathbf{n}}_{p}}$ are independent of each other, and because the random phase ${{\theta }_{k,2n}}$ is introduced, the two noises are not correlated and interference terms from other users using the same pilot can also be eliminated. Therefore, the estimate of ${\mathbf{\Lambda}}_{k}$ can be obtained as
\begin{equation}\label{cov_es_1}
{{{\hat{\mathbf{\Lambda} }}_{k}}=\frac{1}{{{N}_\mathbf{\Lambda}}\rho }\sum\limits_{n=1}^{{{N}_\mathbf{\Lambda}}}{\mathbf{\hat{h}}_{k}^{(1)}[2n-1]{{\left( \mathbf{\hat{h}}_{k}^{(2)}[2n] \right)}^\text{H}}}}.
\end{equation}

However, for a finite ${{N}_\mathbf{\Lambda}}$ , the estimate ${{\hat{\mathbf{\Lambda} }}_{k}}$ in (\ref{cov_es_1}) is not necessarily Hermitian symmetric. Therefore, this matrix can be regularized by approximating it with a positive semidefinite matrix. We approximate ${{\hat{\mathbf{\Lambda} }}_{k}}$ with the positive semidefinite matrix closest to the Frobenius norm, i.e.,
$\hat{\mathbf{\Lambda}}_{k}^\text{PSD}\triangleq \mathbf{U}{{\mathbf{D}}_{+}}{{\mathbf{U}}^{\text{H}}}$, where ${{\mathbf{D}}_{+}}$ is a diagonal matrix that contains only the positive eigenvalues of the symmetric part of ${{\hat{\mathbf{\Lambda} }}_{k}}$, i.e., ${{\mathbf{D}}_{+}}\triangleq {\left( {{{\hat{\mathbf{\Lambda}}}}_{k}}+\hat{\mathbf{\Lambda}}_{k}^{\text{H}} \right)}/{2}\;$, and $\mathbf{U}$ contains the corresponding eigenvectors. Therefore, $\hat{\mathbf{\Lambda}}_k$ is written in the form of $\hat{\mathbf{\Lambda}}_{k}^\text{PSD}$, which is written as
\begin{align}\label{lambda_hat}
{{\hat{\mathbf{\Lambda} }}_{k}}=&\frac{1}{2{{N}_\mathbf{\Lambda}}\rho }\sum\limits_{n=1}^{{{N}_\mathbf{\Lambda}}}({\mathbf{\hat{h}}_{k}^{(1)}[2n-1]{{\left( \mathbf{\hat{h}}_{k}^{(2)}[2n] \right)}^\text{H}}}\nonumber\\
&+{{{\left( \mathbf{\hat{h}}_{k}^{(1)}[2n-1] \right)}^\text{H}}\mathbf{\hat{h}}_{k}^{(2)}[2n]}),
\end{align}
where ${{N}_\mathbf{\Lambda}}$ represents the number of coherent block pairs required to estimate ${\mathbf{\Lambda}}_{k}$. When ${{N}_\mathbf{\Lambda}}\to \infty $, the estimated covariance matrix converges to the real covariance matrix, i.e., ${{\hat{\mathbf{\Lambda }}}_{k}}\to {\mathbf{\Lambda }_{k}}$.

In summary, we substitute the estimated ${{\hat{\mathbf{\Lambda }}}_{k}}$ and ${{\hat{\mathbf{\Sigma }}}_{p}}$ into (\ref{channel_es}), and we can get the channel estimation when the covariance matrix is imperfect
\begin{equation}
{{\mathbf{\hat{g}}}_{k}}={{\hat{\mathbf{\Lambda }}}_{k}}\hat{\mathbf{\Sigma }}_{p}^{-1}{{\mathbf{y}}_{p}}={{\mathbf{\hat{W}}}_{k}}{{\mathbf{y}}_{p}},
\end{equation}
where
\begin{equation}
{{\mathbf{\hat{W}}}_{k}}={{\hat{\mathbf{\Lambda }}}_{k}}\hat{\mathbf{\Sigma }}_{p}^{-1}
\end{equation}
is the deterministic matrix for the estimators.

\begin{remark}
We select alternate pilots with different phase shifts for individual covariance matrix estimation, and then use interval phase-shift-free pilots for sample covariance estimation.  In this scheme, we can obtain accurate covariance matrix estimation and accurate channel estimation without adding additional pilots in the case of low system complexity.
\end{remark}

\section{Uplink Performance with Imperfect Channel Covariance Information}
In this section, we analyze the spectral efficiency of the system based on the above channel estimation method. To derive the closed-form expressions of the achievable rates, we first give the ideal lower boundary of the user achievable rates and then use random matrix theory to perform expectation transformation.

The signal model (\ref{equ_channel_r}) received by the APs can be rewritten as\cite{marzetta2016fundamentals}
\begin{align}\label{wy}
\mathbf{w}_{k}^\text{H}{{\mathbf{y}}_{p}}=&\sqrt{\rho }\mathbb{E}\left[ \mathbf{w}_{k}^\text{H}{{\mathbf{g}}_{k}} \right]+\sqrt{\rho }\left( \mathbf{w}_{k}^\text{H}{{\mathbf{g}}_{k}}-\mathbb{E}\left[ \mathbf{w}_{k}^\text{H}{{\mathbf{g}}_{k}} \right] \right)\nonumber\\
&+\sum\limits_{i\ne k}{\sqrt{\rho }\mathbf{w}_{k}^\text{H}{{\mathbf{g}}_{i}}}+\mathbf{w}_{k}^\text{H}{{\mathbf{n}}_{p}},
\end{align}
where ${{\mathbf{w}}_{k}}$ represents the receiving vector, which can be defined as
\begin{equation}
{{\mathbf{w}}_{k}}= \left\{ \begin{aligned}
&{{{\mathbf{\hat{g}}}}_{k}}, &&\text{for MRC},\\
&{{{\mathbf{\hat{z}}}}_{k}}, &&\text{for ZF},
\end{aligned} \right.
\end{equation}
where ${{{\mathbf{\hat{z}}}}_{k}}$ is the $k$th column of $\mathbf{\hat{G}}{{\left( {{{\mathbf{\hat{G}}}}^{\text{H}}}\mathbf{\hat{G}} \right)}^{-1}}$, $\mathbf{\hat{G}}=\left[ {{{\mathbf{\hat{g}}}}_{1}},{{{\mathbf{\hat{g}}}}_{2}},\cdots ,{{{\mathbf{\hat{g}}}}_{K}} \right]$, and ${{\mathbf{\hat{g}}}_{k}}$ is the estimation of ${{\mathbf{g}}_{k}}$.

The lower boundary of the user achievable rate can be expressed as\cite{upadhya2018covariance}
\begin{equation}
{{R}_{k}}=\left( 1-\frac{P}{{{\tau }_{c}}} \right){{\log }_{2}}\left( 1+{{\gamma }_{k}} \right),
\end{equation}
where $P/{{\tau }_{c}}$ represents the symbols occupied by the pilot. According to (\ref{wy}), the signal-to-noise ratio (SINR) ${{\gamma }_{k}}$ is defined as
\begin{equation}\label{s_n_r}
{{\gamma }_{k}}{\rm{ = }}\frac{{{\left| \mathbb{E}\left[ \mathbf{w}_{k}^{\operatorname{H}}{{\mathbf{g}}_{k}} \right] \right|}^{2}}}{\sum\limits_{i=1}^{K}{\mathbb{E}\left[ {{\left| \mathbf{w}_{k}^{\operatorname{H}}{{\mathbf{g}}_{i}} \right|}^{2}} \right]}-{{\left| \mathbb{E}\left[ \mathbf{w}_{k}^{\operatorname{H}}{{\mathbf{g}}_{k}} \right] \right|}^{2}}+{{\sigma }^{2}}\mathbb{E}\left[ \mathbf{w}_{k}^{\operatorname{H}}{{\mathbf{w}}_{k}} \right]}.
\end{equation}

\subsection{MRC}
When MRC receiver is used, the receiver vector is equal to the channel estimate. Substituting (\ref{channel_es}) into (\ref{s_n_r}) and based on Lemma 1 shown below, we can get the closed-form expression of uplink achievable rates.
\begin{lemma}
	For a complex Wishart matrix $\mathbf{W}$ satisfies ${{\mathcal{W}}}(n,\mathbf{I})$ and $n>m$, it has the property\cite{tulino2004random}
	\begin{equation}\label{w_1}
	\mathbb{E}\left[ \text{tr}\left\{ {{\mathbf{W}}^{-1}} \right\} \right]=\frac{m}{n-m}\\.
	\end{equation}
	For $n>m+1$
	\begin{align}
	\label{w_2}
	\mathbb{E}\left[ \text{tr}\left\{ {{\mathbf{W}}^{-2}} \right\} \right]&=\frac{mn}{{{\left( n-m \right)}^{3}}-\left( n-m \right)},\\
	\label{w_3}
	\mathbb{E}\left[ {{\left| \text{tr}\left( {{\mathbf{W}}^{-1}}\mathbf{A} \right) \right|}^{2}} \right]&=\frac{{{\left| \text{tr}\left( \mathbf{A} \right) \right|}^{2}}+\frac{1}{n-m}\text{tr}\left( \mathbf{A}{{\mathbf{A}}^{H}} \right)}{{{\left( n-m \right)}^{2}}-1},
	\end{align}
	where  $\mathbf{A}\in {{\mathbb{C}}^{m\times m}}$ is a random matrix.
\end{lemma}

\begin{theorem}\label{thm_1}
When MRC receiver is used, the closed-form approximation of SINR is given by
	\begin{equation}
	\gamma _{k}^{\text{MRC}}=\frac{\frac{{{N}_\mathbf{\Sigma }}}{{{N}_\mathbf{\Sigma }}-M\times N}\operatorname{tr}( \mathbf{\bar{W}}_{k}^{H}{\mathbf{\Lambda }_{k}} )}{\sum\limits_{i=1}^{K}{\mathcal{I}_{i}^{\text{EX}}}+\sum\limits_{i\in {{\mathcal{U}}_{p}}}{\mathcal{I}_{i}^{\text{IN}}}-{\frac{{{N}_\mathbf{\Sigma }}}{{{N}_\mathbf{\Sigma }}-M\times N}\operatorname{tr}( \mathbf{\bar{W}}_{k}^{H}{\mathbf{\Lambda }_{k}} )}+{{\sigma }^{2}}{{\mathcal{N}}_{k}}},
	\end{equation}
where
${\mathcal{I}_{i}^{\text{EX}}}
=\frac{{{\mu }_{1}}MN}{2{{N}_\mathbf{\Lambda }}}$ $\operatorname{tr}( {\mathbf{\Lambda }_{i}}{\mathbf{\Sigma }_{p}} )
+\frac{{{\mu }_{1}}}{2{{N}_\mathbf{\Lambda }}}\operatorname{tr}( \mathbf{W}_{k}^{\operatorname{H}} )\operatorname{tr}( {\mathbf{\Lambda }_{i}}{\mathbf{\Lambda }_{k}} )+{{\mu }_{1}}\operatorname{tr}( \mathbf{W}_{k}^{\operatorname{H}}{\mathbf{\Lambda }_{i}}{\mathbf{\Lambda }_{k}} )$,
	${\mathcal{I}_{i}^{\text{IN}}}
=\frac{{{\mu }_{1}}}{2{{N}_\mathbf{\Sigma }}{{N}_\mathbf{\Lambda }}}\operatorname{tr}(\mathbf{\Sigma} _{p}^{-2}{\mathbf{\Lambda }_{k}} )\operatorname{tr}( {\mathbf{\Lambda }_{i}}^{2}{\mathbf{\Lambda }_{k}} ) +\frac{{{\mu }_{1}}}{{{N}_\mathbf{\Sigma }}}\operatorname{tr}( \mathbf{W}_{k}^{\operatorname{H}}{\mathbf{\Lambda }_{i}}^{2}{{\mathbf{W}}_{k}} ) +\frac{{{\mu }_{1}}MN}{2{{N}_\mathbf{\Sigma }}{{N}_\mathbf{\Lambda }}}\operatorname{tr}( \mathbf{\Sigma} _{p}^{-1} )\operatorname{tr}( {\mathbf{\Lambda }_{i}}^{2}{\mathbf{\Sigma }_{p}} )+ {{\mu }_{2}}{{| \operatorname{tr}( {\mathbf{\Lambda }_{k}}{{\mathbf{W}}_{i}} ) |}^{2}}	+\frac{{{\mu }_{2}}}{2{{N}_\mathbf{\Lambda }}} \operatorname{tr}( {{\mathbf{W}}_{i}}{\mathbf{\Sigma }_{p}}\mathbf{W}_{i}^{\operatorname{H}}{\mathbf{\Sigma }_{p}})
 +\frac{{{\mu }_{2}}}{2{{N}_\mathbf{\Lambda }}} $ $ \operatorname{tr}( {{\mathbf{W}}_{i}}{\mathbf{\Lambda }_{k}}\mathbf{W}_{i}^{\operatorname{H}}{\mathbf{\Lambda }_{k}} )
$, ${{\mathcal{N}}_{k}}
={{\mu }_{1}}\operatorname{tr}( {{\mathbf{W}}_{k}}{\mathbf{\Lambda }_{k}} )+\frac{{{\mu }_{1}}MN}{2{{N}_\mathbf{\Lambda }}}\operatorname{tr}( {\mathbf{\Sigma }_{p}})+\frac{{{\mu }_{1}}}{2{{N}_\mathbf{\Lambda }}}\operatorname{tr}( {\mathbf{\Lambda }_{k}} )\operatorname{tr}( \mathbf{W}_{k}^{\operatorname{H}} )$,
	${{\mathbf{\bar{W}}}_{k}}={{\bar{\mathbf{\Lambda }}}_{k}}\mathbf{\Sigma} _{p}^{-1}$,
	${{\bar{\mathbf{\Lambda} }}_{k}}=\mathbb{E}\left[ {{{\hat{\mathbf{\Lambda} }}}_{k}} \right]={{\mathbf{\Lambda} }_{k}}$,
	${{\mu }_{1}}=\frac{{{N}_\mathbf{\Sigma }}^{3}}{\left[ {{\left( {{N}_\mathbf{\Sigma }}-M\times N \right)}^{2}}-1 \right]\left( {{N}_\mathbf{\Sigma }}-M\times N \right)}$ and
	${{\mu }_{2}}=\frac{{{N}_\mathbf{\Sigma }}^{2}}{{{\left( {{N}_\mathbf{\Sigma }}-M\times N \right)}^{2}}-1}$.
	{\emph{Proof: }}Please refer to Appendix \ref{ap}.
\end{theorem}

In addition, when the total number of antennas $MN\to \infty $, the limit rate of each user given in (\ref{long}) at the top of next page.
\begin{figure*}[!t]
	\normalsize
	\setcounter{mytempeqncnt}{\value{equation}}
	\setcounter{equation}{26}
	\begin{equation}\label{long}
	\gamma _{k}^{\text{MRC}}\to\frac{{\left| \operatorname{tr}\left( \mathbf{\bar{W}}_{k}^{H}{\mathbf{\Lambda }_{k}} \right) \right|}^{2}}
	{\begin{aligned}
		&\sum\limits_{i\in {{\mathcal{U}}_{p}}}{\left( {{\left| \operatorname{tr}\left( {\mathbf{\Lambda }_{k}}{\mathbf{\Lambda }_{i}}{\mathbf{\Sigma }_{p}} \right) \right|}^{2}}+\frac{1}{2{{N}_\mathbf{\Lambda }}}\left( \operatorname{tr}\left( {\mathbf{\Lambda }_{i}}{\mathbf{\Sigma }_{p}}{{\left( {\mathbf{\Lambda }_{i}}{\mathbf{\Sigma }_{p}} \right)}^{\text{H}}}\left( \mathbf{\Sigma} _{p}^{2}+\mathbf{\Lambda} _{k}^{2} \right) \right)+\operatorname{tr}\left( \mathbf{\Sigma} _{p}^{-1} \right)\operatorname{tr}\left( {\mathbf{\Lambda }_{i}}^{2}{\mathbf{\Sigma }_{p}} \right) \right) \right)} \\		
		+&\sum\limits_{i=1}^{K}{\frac{{{N}_\mathbf{\Sigma }}}{2{{N}_\mathbf{\Lambda }}}\operatorname{tr}\left( {\mathbf{\Lambda }_{i}}{\mathbf{\Sigma }_{p}} \right)}+\frac{1}{\rho }\frac{{{N}_\mathbf{\Sigma }}}{2{{N}_\mathbf{\Lambda }}}\operatorname{tr}\left( {\mathbf{\Sigma }_{p}} \right)-{{\left| \operatorname{tr}\left( \mathbf{\bar{W}}_{k}^{H}{\mathbf{\Lambda }_{k}} \right) \right|}^{2}}\\
		\end{aligned}}
	\end{equation}
	
	\hrulefill
	\vspace*{4pt}
\end{figure*}
It can be seen that when the MRC receiver is used, even if the AP is equipped with a large number of antennas, the SINR of the receiver still tends to be a constant related to the covariance matrix, and its performance still has a bottleneck. It can also be seen that the SINR of the receiver can be improved by reasonably allocating users using the same pilot.

\subsection{ZF}
After ZF detection, the received signal can be written as
\begin{equation}
{{\left( {{{\mathbf{\hat{G}}}}^{\text{H}}}\mathbf{\hat{G}} \right)}^{-1}}{{\mathbf{\hat{G}}}^{\text{H}}}\mathbf{y}=\mathbf{x}+{{\left( {{{\mathbf{\hat{G}}}}^{\text{H}}}\mathbf{\hat{G}} \right)}^{-1}}{{\mathbf{\hat{G}}}^{\text{H}}}\left( \mathbf{\tilde{G}x}+\mathbf{n} \right),
\end{equation}
where $\mathbf{\tilde{G}}=\left[ {{{\mathbf{\tilde{g}}}}_{1}},{{{\mathbf{\tilde{g}}}}_{2}},\cdots ,{{{\mathbf{\tilde{g}}}}_{K}} \right]$, ${{\mathbf{\tilde{g}}}_{k}}$ represents the error of channel estimation
\begin{equation}
{{\mathbf{\tilde{g}}}_{k}}={{\mathbf{g}}_{k}}-{{\mathbf{\hat{g}}}_{k}}.
\end{equation}

Therefore, the SINR ${{\gamma }_{k}}$ can be expressed as
\begin{equation}\label{sinr_zf}
\gamma _{k}^{\text{ZF}}=\frac{1}{\mathbf{e}_{k}^{\text{H}}{{\left( {{{\mathbf{\hat{G}}}}^{\text{H}}}\mathbf{\hat{G}} \right)}^{-1}}{{\mathbf{\hat{G}}}^{\text{H}}}\tilde{\mathbf{\Gamma }}{{\mathbf{\hat{G}}}^{\text{H}}}{{\left( {{{\mathbf{\hat{G}}}}^{\text{H}}}\mathbf{\hat{G}} \right)}^{-1}}{{\mathbf{e}}_{k}}},
\end{equation}
where $\tilde{\mathbf{\Gamma }}$ is defined as the interference item
\begin{align}
\tilde{\mathbf{\Gamma}}=&\operatorname{cov}\left( \mathbf{\tilde{G}x}+\mathbf{n} \right) \nonumber\\
=&\sum\limits_{k=1}^{K}{\left( {\mathbf{\Lambda }_{k}}-{{{\hat{\mathbf{\Lambda }}}}_{k}}\hat{\mathbf{\Sigma}}_{p}^{-2}{{{\hat{\mathbf{\Lambda }}}}_{k}}{\mathbf{\Sigma }_{p}} \right)}+{\mathbf{\sigma }^{2}}{{\mathbf{I}}_{MN}}.
\end{align}
\begin{lemma}
	For $k\in {{\mathcal{U}}_{p}}$, the channels between the users using the same pilot and the APs have the following relationship
	\begin{align}
	\frac{1}{MN}\mathbf{\hat{G}}_{p}^{\text{H}}{{\mathbf{\hat{G}}}_{i}}&\xrightarrow[MN\to \infty ]{a.s.}0,p\ne i,\\
	\frac{1}{MN}\mathbf{\hat{G}}_{p}^{\text{H}}\tilde{\mathbf{\Gamma}}{{\mathbf{\hat{G}}}_{i}}&\xrightarrow[MN\to \infty ]{a.s.}0,p\ne i,\\
	\frac{1}{MN}\mathbf{\hat{G}}_{p}^{\text{H}}{{\mathbf{\hat{G}}}_{p}}-\frac{1}{M}{\mathbf{\Xi }_{p}}&\xrightarrow[MN\to \infty ]{a.s.}0,\\
	\frac{1}{MN}\mathbf{\hat{G}}_{p}^{\text{H}}\tilde{\mathbf{\Gamma}}{{\mathbf{\hat{G}}}_{p}}-\frac{1}{M}{{\tilde{\mathbf{\Xi }}}_{p}}&\xrightarrow[MN\to \infty ]{a.s.}0,
	\end{align}
	where, for $i\in {{\mathcal{U}}_{p}}$ has
	\begin{align}
	{{\left[ {\mathbf{\Xi }_{p}} \right]}_{q,j}}&=\operatorname{tr}\left( {{{\hat{\mathbf{\Lambda}}}}_{k}}\hat{\mathbf{\Sigma }}_{p}^{-1}\hat{\mathbf{\Sigma}}_{p}^{-1}{{{\hat{\mathbf{\Lambda }}}}_{i}}{\mathbf{\Sigma}_{p}} \right),\\
	{{\left[ {{{\tilde{\mathbf{\Xi}}}}_{p}} \right]}_{q,j}}&=\operatorname{tr}\left( {{{\hat{\mathbf{\Lambda}}}}_{k}}\hat{\mathbf{\Sigma}}_{p}^{-1}\hat{\mathbf{\Sigma}}_{p}^{-1}{{{\hat{\mathbf{\Lambda }}}}_{i}}{\mathbf{\Sigma}_{p}}\tilde{\mathbf{\Gamma}}\right).
	\end{align}
\end{lemma}

Based on the above expressions, when $MN\to \infty$, (\ref{sinr_zf}) can be transformed into
\begin{equation}\label{sinr_zf_1}
{{\gamma }_{k}^{\text{ZF}}}=\frac{N}{\mathbf{e}_{q}^{\text{H}}\mathbf{\Xi}_{p}^{-1}{{{\tilde{\mathbf{\Xi}}}}_{p}}\mathbf{\Xi}_{p}^{-1}{{\mathbf{e}}_{q}}}.
\end{equation}

Next, by approximating the estimates, the theoretical expression of SINR of the ZF receiver can be obtained.
\begin{theorem}
	When ZF receiver is used, the closed-form approximation of SINR is given by
	\begin{equation}\label{r_zf}
	\gamma _{k}^{\text{ZF}}=\frac{N}{\mathbf{e}_{q}^{\text{H}}{{\left( \mathbf{\Xi }_{p}^{\text{th}} \right)}^{-1}}\tilde{\mathbf{\Xi } }_{p}^{\text{th}}{{\left(\mathbf{\Xi }_{p}^{\text{th}} \right)}^{-1}}{{\mathbf{e}}_{q}}},
	\end{equation}
	where
	\begin{align}
	{{\left[ {\mathbf{\Xi }_{p}^{\text{th}}} \right]}_{q,j}}
	=&{{\mu }_{1}}\operatorname{tr}\left( {\mathbf{\Lambda }_{k}}\mathbf{\Sigma} _{p}^{-1}{\mathbf{\Lambda }_{i}} \right)+\frac{{{\mu }_{1}}MN}{2{{N}_\mathbf{\Lambda }}}\operatorname{tr}\left( {\mathbf{\Sigma }_{p}} \right) \nonumber\\
	& +\frac{{{\mu }_{1}}}{2{{N}_\mathbf{\Lambda }}}\operatorname{tr}\left( {\mathbf{\Lambda }_{k}} \right)\operatorname{tr}\left( \mathbf{\Sigma} _{p}^{-1}{\mathbf{\Lambda }_{i}} \right),\\
	{{\left[ {{{\tilde{\mathbf{\Xi }}}}_{p}^{\text{th}}} \right]}_{q,j}}
	=&{{\mu }_{1}}\text{tr}\left( {\mathbf{\Lambda }_{k}}\mathbf{\Sigma} _{p}^{-1}{\mathbf{\Lambda }_{i}}\tilde{\mathbf{\Gamma}} \right)+\frac{{{\mu }_{1}}MN}{2{{N}_\mathbf{\Lambda }}}\operatorname{tr}\left( {\mathbf{\Sigma }_{p}}\tilde{\mathbf{\Gamma }}\right) \nonumber\\
	&+\frac{{{\mu }_{1}}}{2{{N}_\mathbf{\Lambda }}}\operatorname{tr}\left( {\mathbf{\Lambda }_{k}}\tilde{\mathbf{\Gamma}} \right)\operatorname{tr}\left( \mathbf{\Sigma} _{p}^{-1}{\mathbf{\Lambda }_{i}} \right),
	\end{align}
	and $\tilde{\mathbf{\Gamma}}$ is (\ref{tilde_top}) at the top of the page,
	\begin{figure*}[!t]
		\normalsize
		\setcounter{mytempeqncnt}{\value{equation}}
		\setcounter{equation}{41}
		\begin{equation}\label{tilde_top}
		\tilde{\mathbf{\Gamma }}=\sum\limits_{k=1}^{K}{\left[ {\mathbf{\Lambda }_{k}}-\left(
			N_{\mathbf{\Sigma}}^{2}{\mathbf{\Lambda }_{k}}\tilde{\mathbf{\Sigma }}_{p}^{-1}\mathbf{\Sigma} _{p}^{-1}\tilde{\mathbf{\Sigma}}_{p}^{-1}{\mathbf{\Lambda }_{k}}
			+\frac{{{\mu }_{1}}}{2{{N}_\mathbf{\Lambda }}}\mathbf{\Sigma}_{p}^{2}\operatorname{tr}\left( \mathbf{\Sigma} _{p}^{-1} \right)
			+\frac{{{\mu }_{1}}}{2{{N}_\mathbf{\Lambda }}}{\mathbf{\Lambda }_{k}}{\mathbf{\Sigma }_{p}}\operatorname{tr}\left( {\mathbf{\Lambda }_{k}} \right)
			\right) \right]}+{\mathbf{\sigma }^{2}}{{\mathbf{I}}_{MN}}.
		\end{equation}
		
		\hrulefill
		\vspace*{4pt}
	\end{figure*}
	where ${{\tilde{\mathbf{\Sigma}}}_{p}}$ is a Wishart matrix, and it satisfies $\mathcal{W}\left( {{N}_\mathbf{\Sigma }},{{\mathbf{I}}_{MN}} \right)$.
	
	{\emph{Proof: }}Please refer to Appendix \ref{bp}.
\end{theorem}

It can be seen that unlike MRC, using ZF receiver, when the number of antennas of each APs $N\to \infty $, the SINR increases linearly as (\ref{r_zf}). At the same time, it can be known that the characteristic of matrix ${\mathbf{\Xi }_{p}}$ is an important factor that restricts system performance, and it is closely related to pilot contamination. When ${\mathbf{\Xi }_{p}}$ is rank-deficient, the performance of the ZF receiver will also be seriously degraded.

In summary, the closed-form expressions of SINR in both MRC and ZF receivers are related to the number of coherent blocks $N_{\mathbf{\Sigma}}$ and $N_{\mathbf{\Lambda}}$ to calculate the covariance matrix. Better estimation of the covariance matrix can lead to higher SINR and thus improve the system performance.

\section{Simulation Results}
In this section, we fit the closed-form expressions with the simulation value, and compare them with the case where the covariance matrix is perfect. With the numerical result, we analyze the accuracy of channel estimation with imperfect covariance matrix.

We consider a cell-free distributed massive MIMO system, and there are $M=5$ APs with $N=50$ antennas randomly distributed in the area. We use the channel model in (\ref{ch_model}), where the path loss exponent $\zeta $ is 3.7, the reference distance is 1, and the variance of shadow fading satisfies exponential normal distribution. Set all APs to be uniformly distributed in a circle with a diameter of 1 km, and all $K=5$ users are randomly distributed in the area. We set the number of coherent blocks with constant channel covariance matrix ${{\tau }_{s}}=20000$, the number of symbols in a coherent block ${{\tau }_{c}=200}$ , and the number of pilot symbols ${\tau=10}$.
For comparison, we consider the following three aspects.
\begin{itemize}
	\item \textbf{Channel estimation with perfect covariance matrix(Cov known):} The covariance matrix satisfies (\ref{equ_channelM}), and we perform the MMSE estimation of the channel to get the sum of user rates.
	\item \textbf{The simulated value of channel estimation with imperfect covariance matrix(Simulated):} We directly use the covariance matrix estimation results of (\ref{sigma_hat}) and (\ref{lambda_hat}) for channel estimation, and get the sum of user rates to judge system performance.
	\item \textbf{The theoretical value of channel estimation with imperfect covariance matrix(Theoretical):} We derive the Theorem 1 and the Theorem 2, and obtain the closed expression of the SINR to get the sum of user rates.
\end{itemize}
Based on the above three aspects, we simulate MRC and ZF respectively.
\begin{figure}
	\centering
	\includegraphics[width=9cm]{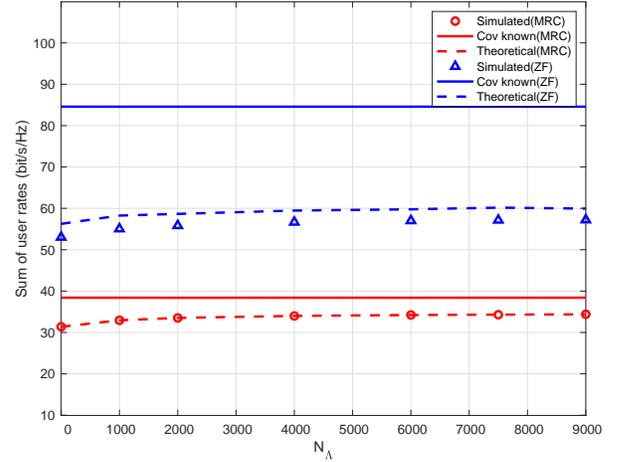}\\
	\caption{The sum of user rates when ${{N}_\mathbf{\Sigma }}=1000$ with completely orthogonal pilots}\label{fig_2}
\end{figure}

\begin{figure}
	\centering
	\includegraphics[width=9cm]{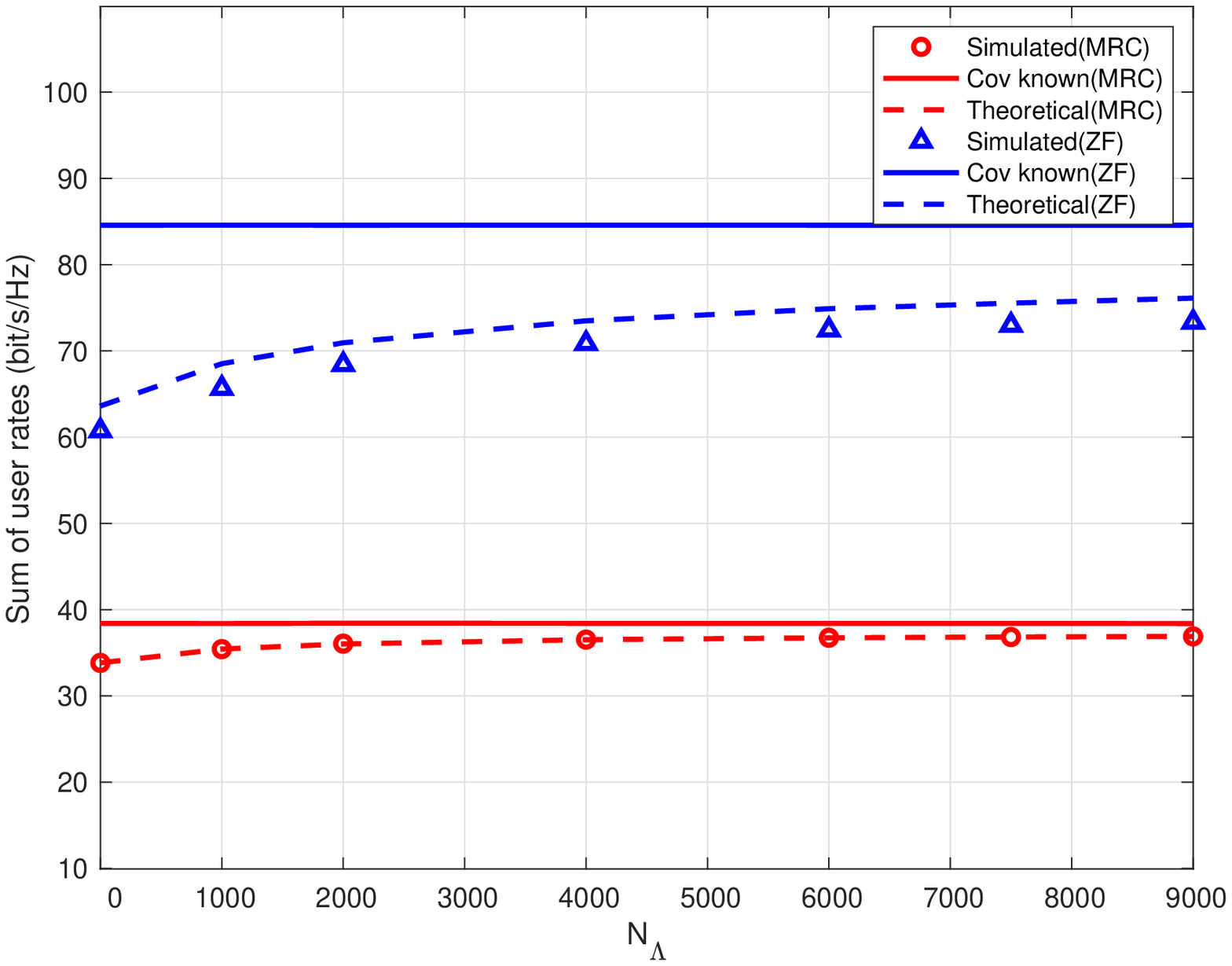}\\
	\caption{The sum of user rates when ${{N}_\mathbf{\Sigma }}=3000$ with completely orthogonal pilots}\label{fig_3}
\end{figure}

\begin{figure}
	\centering
	\includegraphics[width=9cm]{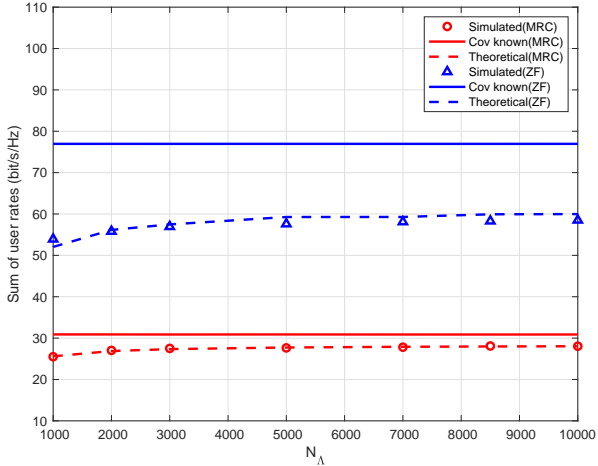}\\
	\caption{The sum of user rates when ${{N}_\mathbf{\Sigma }}=1000$ with P=8 pilots}\label{fig_4}
\end{figure}

Fig. \ref{fig_2} and Fig. \ref{fig_3} respectively show the sum of user rates when ${{N}_\mathbf{\Sigma }}=1000$ and ${{N}_\mathbf{\Sigma }}=3000$ with completely orthogonal pilots. It can be seen from Fig. \ref{fig_2} and Fig. \ref{fig_3} that the simulated value fits well with the theoretical value. Therefore, the correctness of our derived closed-form expressions is proved. When using MRC and ZF, the sum of user rates with perfect covariance are close to the situation when the covariance is imperfect, and MRC is closer than ZF. We can even see that when using MRC receivers, our covariance estimation scheme can achieve 95\% or higher spectral efficiency with a perfect covariance matrix. This proves that the covariance matrix estimation method we used can bring good system performance. But when using ZF, the sum of user rates are higher than using MRC. This is because the ZF receiver can eliminate inter-user interference. From the curve of the sum of user rates with perfect covariance matrix, it can be seen that the sum of user rates increase as the number of pilots ${{N}_\mathbf{\Lambda }}$ used to calculate the individual covariance estimate increases. This result is due to the fact that the covariance estimation is not accurate when the inserted ${{N}_\mathbf{\Lambda }}$ is small, and the inaccurate channel estimation results in a lower sum of user rates. As ${{N}_\mathbf{\Lambda }}$ continues to increase, the accuracy of the channel estimation improves, and the sum of user rates increase accordingly.

In addition, comparing Fig. \ref{fig_2} and Fig. \ref{fig_3}, it can be seen that the number coherent blocks ${{N}_\mathbf{\Sigma }}$ used to estimate the covariance matrix ${\mathbf{\Sigma }_{p}}$ of the received signal will actually affect the system performance, and a more accurate estimation of ${\mathbf{\Sigma }_{p}}$ will have better system performance. The increase of ${{N}_\mathbf{\Sigma }}$ does not change the trend of the curve, because the increase of ${{N}_\mathbf{\Sigma }}$ is only a better estimate of ${\mathbf{\Sigma }_{p}}$ under the existing structure, and does not change the relationship between sum of user rates and ${{N}_\mathbf{\Lambda }}$. At this time, the sum of user rates of the imperfect covariance matrix are closer to the perfect situation. This is at the cost of calculation. In actual situations, a good CPU to increase ${{N}_\mathbf{\Sigma }}$ to get better system performance is required.

Fig. \ref{fig_4} shows the sum of user rates when ${{N}_\mathbf{\Sigma }}=1000$ with $P=8$ pilots. Because of the large number of users in a dense distributed MIMO system, the situation of each user using completely orthogonal pilots cannot be satisfied. So it is necessary to consider the situation of pilot contamination, that is, multiple users use the same pilot. It can be seen that pilot contamination has a certain impact on the performance of MRC and ZF. Pilot multiplexing can improve system resource utilization, and a good pilot allocation method can achieve a trade-off between resource utilization and system performance.

\section{Conclusion}
This paper proposes a method to estimate the covariance matrix by alternating pilot in adjacent coherent blocks. This method improves the channel estimation method without extra pilot overhead to achieve better system performance when the covariance matrix is imperfect. We derived the closed-form expression of uplink SE with MRC and ZF receivers and verified the closed-form expressions. We also analyze the influence of the covariance matrix calculation with different number of coherent blocks on the system performance. The simulation results show that the channel estimation algorithm with imperfect channel covariance information can achieve high estimation accuracy. At the same time, a better estimation of the covariance matrix of the received signal leads to better system performance. In addition, for different receivers, pilot contamination can decrease system performance.

\appendices
\section{Proof of Theorem 1}\label{ap}
Based on the random matrix theory (see (\cite{tulino2004random})), we provide the following proof which is divided into two steps.
First, if the combining vector ${{\mathbf{w}}_{k}}={{\mathbf{\hat{W}}}_{k}}{{\mathbf{y}}_{p}}$ is used, we need to convert the expression to ${{\mathbf{\hat{W}}}_{k}}$. For $\mathbb{E}\left[ \mathbf{w}_{k}^{\text{H}}{{\mathbf{g}}_{k}} \right]$,
\begin{equation}
\mathbb{E}\left[ \mathbf{w}_{k}^{\operatorname{H}}{{\mathbf{g}}_{k}} \right]
=\mathbb{E}\left[ \mathbf{y}_{p}^{\operatorname{H}}\mathbf{\hat{W}}_{k}^{\operatorname{H}}{{\mathbf{g}}_{k}} \right]
\overset{\text{(a)}}{=}{{\mathbb{E}}_{\mathbf{W}}}\left[ \operatorname{tr}\left( \mathbf{\hat{W}}_{k}^{\operatorname{H}}{\mathbf{\Lambda }_{k}} \right) \right],
\end{equation}
where (a) is obtained because the noise ${{\mathbf{n}}_{p}}$ and the channel ${{\mathbf{g}}_{k}}$ are independent of each other, and ${{\mathbf{x}^\text{H}}\mathbf{y}=\text{tr}\left(\mathbf{y}{\mathbf{x}^\text{H}}\right)}$ for any vectors $\mathbf{x}$, $\mathbf{y}$.
By using the same trace rule as above and identifying $\mathbf{\Sigma}_p$, for $\mathbb{E}\left[ \mathbf{w}_{k}^{\text{H}}{{\mathbf{w}}_{k}} \right]$, we have
\begin{align}
\mathbb{E}\left[ \mathbf{w}_{k}^{\operatorname{H}}{{\mathbf{w}}_{k}} \right]
=&\mathbb{E}\left[ {{\left( {{\mathbf{y}}_{p}} \right)}^{\operatorname{H}}}\mathbf{\hat{W}}_{k}^{\operatorname{H}}{{{\mathbf{\hat{W}}}}_{k}}{{\mathbf{y}}_{p}} \right] \nonumber\\
=&{{\mathbb{E}}_{\operatorname{W}}}\left[ \operatorname{tr}\left( {{{\mathbf{\hat{W}}}}_{k}}\mathbb{E}\left[ {{\mathbf{y}}_{p}}{{\left( {{\mathbf{y}}_{p}} \right)}^{\operatorname{H}}} \right]\mathbf{\hat{W}}_{k}^{\operatorname{H}} \right) \right]\nonumber\\
=&{{\mathbb{E}}_{\operatorname{W}}}\left[ \operatorname{tr}\left( {{{\mathbf{\hat{W}}}}_{k}}{\mathbf{\Sigma }_{p}}\mathbf{\hat{W}}_{k}^{\operatorname{H}} \right) \right].
\end{align}

Finally, for $\mathbb{E}\left[ {{\left| \mathbf{w}_{k}^{\operatorname{H}}{{\mathbf{g}}_{i}} \right|}^{2}} \right]$,
\begin{equation}
\mathbb{E}\left[ {{\left| \mathbf{w}_{k}^{\operatorname{H}}{{\mathbf{g}}_{i}} \right|}^{2}} \right]=\mathbb{E}\left[ \mathbf{w}_{k}^{\operatorname{H}}{{\mathbf{g}}_{i}}\mathbf{g}_{i}^{\operatorname{H}}{{\mathbf{w}}_{k}} \right].
\end{equation}
If $i\notin {{\mathcal{U}}_{p}}$, then
\begin{align}
\mathbb{E}\left[ \mathbf{w}_{k}^{\operatorname{H}}{{\mathbf{g}}_{i}}\mathbf{g}_{i}^{\operatorname{H}}{{\mathbf{w}}_{k}} \right]
=&\mathbb{E}\left[ \mathbf{w}_{k}^{\operatorname{H}}{\mathbf{\Lambda }_{i}}{{\mathbf{w}}_{k}} \right] \nonumber\\
=&{{\mathbb{E}}_{\operatorname{W}}}\left[ \operatorname{tr}({{{\mathbf{\hat{W}}}}_{k}}\mathbb{E}\left[ {{\mathbf{y}}_{p}}\mathbf{y}_{p}^{\operatorname{H}} \right]\mathbf{\hat{W}}_{k}^{\operatorname{H}}{\mathbf{\Lambda }_{i}}) \right]\nonumber\\
=&{{\mathbb{E}}_{\operatorname{W}}}\left[ \operatorname{tr}\left( {{{\mathbf{\hat{W}}}}_{k}}{\mathbf{\Sigma }_{p}}\mathbf{W}_{k}^{\operatorname{H}}{\mathbf{\Lambda }_{i}} \right) \right].
\end{align}
If $i\in {{\mathcal{U}}_{p}}$, ${{\mathbf{w}}_{k}}$ and ${{\mathbf{g}}_{i}}$ are not independent of each other, they will be coupled together, and thus cannot be solved according to the above method. So, we have
\begin{align}
&\mathbb{E}\left[ \mathbf{w}_{k}^{\operatorname{H}}{{\mathbf{g}}_{i}}\mathbf{g}_{i}^{\operatorname{H}}{{\mathbf{w}}_{k}} \right] \nonumber\\
\overset{\text{(a)}}{=}&\mathbb{E}\left[ {{\left| {{\left( {{\mathbf{w}}_{k}}-{{{\mathbf{\hat{W}}}}_{k}}{{\mathbf{g}}_{i}} \right)}^{\operatorname{H}}}{{\mathbf{g}}_{i}} \right|}^{2}} \right]+\mathbb{E}\left[ {{\left| \mathbf{g}_{i}^{\operatorname{H}}{{{\mathbf{\hat{W}}}}_{k}}{{\mathbf{g}}_{i}} \right|}^{2}} \right]\nonumber \\
=&{{\mathbb{E}}_{\operatorname{W}}}\left[ \operatorname{tr}({{{\mathbf{\hat{W}}}}_{k}}\left( {\mathbf{\Sigma }_{p}}-{\mathbf{\Lambda }_{i}} \right)\mathbf{\hat{W}}_{k}^{\operatorname{H}}{\mathbf{\Lambda }_{i}})+{{\left| \operatorname{tr}\left( \mathbf{\hat{W}}_{k}^{\operatorname{H}}{\mathbf{\Lambda }_{i}} \right) \right|}^{2}} \right]\nonumber \\
&+{{\mathbb{E}}_{\operatorname{W}}}\left[ \operatorname{tr}\left( {{{\mathbf{\hat{W}}}}_{k}}{\mathbf{\Lambda }_{i}}\mathbf{\hat{W}}_{k}^{\operatorname{H}}{\mathbf{\Lambda }_{i}} \right) \right]\nonumber \\
=&{{\mathbb{E}}_{\operatorname{W}}}\left[ \operatorname{tr}\left( {{{\mathbf{\hat{W}}}}_{k}}{\mathbf{\Sigma }_{p}}\mathbf{\hat{W}}_{k}^{\operatorname{H}}{\mathbf{\Lambda }_{i}} \right) \right]+{{\mathbb{E}}_{\operatorname{W}}}\left[ {{\left| \operatorname{tr}\left( \mathbf{\hat{W}}_{k}^{\operatorname{H}}{\mathbf{\Lambda }_{i}} \right) \right|}^{2}} \right],
\end{align}
where (a) is because ${{\mathbf{w}}_{k}}-{{{\mathbf{\hat{W}}}}_{k}}{{\mathbf{g}}_{i}}$ and ${{\mathbf{g}}_{i}}$ are independent.

Next, continue to solve the expectation of ${{\mathbf{\hat{W}}}_{k}}$ and convert the above expressions into closed-form expressions.
For any random vector $\mathbf{g}$ that is independent of each other and each elements obeying the ${{\cal{N}}_{\rm{c}}}\left( {0,1} \right)$ distribution
\begin{equation}\label{a_1}
\mathbb{E}\left[ \mathbf{g}{{\mathbf{g}}^{\operatorname{H}}}\mathbf{Ag}{{\mathbf{g}}^{\operatorname{H}}} \right]=\mathbf{A}+\mathbf{I}\text{tr}\left( \mathbf{A} \right),
\end{equation}
define
\begin{equation}\label{s_1}
{{\tilde{\mathbf{\Sigma}}}_{p}}={{N}_\mathbf{\Sigma }}\left( \mathbf{\Sigma} _{p}^{-1/2}{{{\hat{\mathbf{\Sigma }}}}_{p}}\mathbf{\Sigma} _{p}^{-1/2} \right),
\end{equation}
then ${{\tilde{\mathbf{\Sigma}}}_{p}}$ is a Wishart matrix, and it satisfies $\mathcal{W}\left( {{N}_\mathbf{\Sigma }},\mathbf{I} \right)$. Hence, for ${{\mathbb{E}}_{\mathbf{W}}}\left[ \operatorname{tr}\left( \mathbf{\hat{W}}_{k}^{\operatorname{H}}{\mathbf{\Lambda }_{k}} \right) \right]$, we have that
\begin{align}
{{\mathbb{E}}_{\mathbf{W}}}\left[ \operatorname{tr}\left( \mathbf{\hat{W}}_{k}^{\operatorname{H}}{\mathbf{\Lambda }_{k}} \right) \right]=&{{\mathbb{E}}_{\mathbf{W}}}\left[ \operatorname{tr}\left( \hat{\mathbf{\Sigma }}_{p}^{-1}{{{\hat{\mathbf{\Lambda} }}}_{k}}{\mathbf{\Lambda }_{k}} \right) \right]\nonumber \\
\overset{\text{(a)}}{=}&{{\mathbb{E}}_\mathbf{\Sigma }}\left[ \operatorname{tr}\left( \mathbf{\Sigma} _{p}^{-1/2}{{N}_\mathbf{\Sigma }}\tilde{\mathbf{\Sigma }}_{p}^{-1}\mathbf{\Sigma} _{p}^{-1/2}{{{\hat{\mathbf{\Lambda }}}}_{k}}{\mathbf{\Lambda }_{k}} \right) \right]\nonumber \\
\overset{\text{(b)}}{=}&{{N}_\mathbf{\Sigma }}{{\mathbb{E}}_\mathbf{\Lambda }}\left[ \operatorname{tr}\left( \frac{\mathbf{\Sigma} _{p}^{-1/2}\mathbf{I}\mathbf{\Sigma} _{p}^{-1/2}}{{{N}_\mathbf{\Sigma }}-M\times N}{{{\hat{\mathbf{\Lambda} }}}_{k}}{\mathbf{\Lambda }_{k}} \right) \right] \nonumber\\
=&\frac{{{N}_\mathbf{\Sigma }}}{{{N}_\mathbf{\Sigma }}-M\times N}\operatorname{tr}\left( \mathbf{\bar{W}}_{k}^{\operatorname{H}}{\mathbf{\Lambda }_{k}} \right),
\end{align}
where (a) and (b) are obtained by substituting (\ref{s_1}) and
(\ref{w_1}), respectively.
In addition, for ${{\mathbb{E}}_{\mathbf{W}}}\left[ \operatorname{tr}\left( {{{\mathbf{\hat{W}}}}_{k}}{\mathbf{\Sigma }_{p}}\mathbf{\hat{W}}_{k}^{\operatorname{H}} \right) \right]$,
\begin{equation}
\begin{aligned}
&{{\mathbb{E}}_{\mathbf{W}}}\left[ \operatorname{tr}\left( {{{\mathbf{\hat{W}}}}_{k}}{\mathbf{\Sigma }_{p}}\mathbf{\hat{W}}_{k}^{\operatorname{H}} \right) \right] \\
= &{{\mathbb{E}}_{\mathbf{W}}}\left[ \operatorname{tr}\left( {{{\hat{\mathbf{\Lambda }}}}_{k}}\hat{\mathbf{\Sigma}}_{p}^{-1}{\mathbf{\Sigma }_{p}}\hat{\mathbf{\Sigma} }_{p}^{-1}{{{\hat{\mathbf{\Lambda }}}}_{k}} \right) \right] \\
\overset{\text{(a)}}{=}&{{N}_\mathbf{\Sigma }}^{2}{{\mathbb{E}}_{\mathbf{W}}}\left[ \operatorname{tr}\left( {{{\hat{\mathbf{\Lambda }}}}_{k}}\mathbf{\Sigma} _{p}^{-1/2}\tilde{\mathbf{\Sigma }}_{p}^{-1}\tilde{\mathbf{\Sigma}}_{p}^{-1}\mathbf{\Sigma} _{p}^{-1/2}{{{\hat{\mathbf{\Lambda }}}}_{k}} \right) \right] \\
\overset{\text{(b)}}{=}&{{\mu }_{1}}{{\mathbb{E}}_\mathbf{\Lambda }}\left[ \operatorname{tr}\left( {{{\hat{\mathbf{\Lambda }}}}_{k}}\mathbf{\Sigma} _{p}^{-1}{{{\hat{\mathbf{\Lambda }}}}_{k}} \right) \right] \\
\overset{\text{(c)}}{=}&{{\mu }_{1}}\operatorname{tr}\left( {{\mathbf{W}}_{k}}{\mathbf{\Lambda }_{k}} \right)+\frac{{{\mu }_{1}}MN}{2{{N}_\mathbf{\Lambda }}}\operatorname{tr}\left( {\mathbf{\Sigma }_{p}} \right)+\frac{{{\mu }_{1}}}{2{{N}_\mathbf{\Lambda }}}\operatorname{tr}\left( {\mathbf{\Lambda }_{k}} \right)\operatorname{tr}\left( \mathbf{W}_{k}^{\operatorname{H}} \right), \\
\end{aligned}
\end{equation}
where (a) , (b) and (c) result from (\ref{s_1}) ,
(\ref{w_2}) and (\ref{a_1}), respectively.
And for ${{\mathbb{E}}_{\mathbf{W}}}\left[ \operatorname{tr}\left( {{{\mathbf{\hat{W}}}}_{k}}{\mathbf{\Sigma }_{p}}\mathbf{\hat{W}}_{k}^{\operatorname{H}}{\mathbf{\Lambda }_{i}} \right) \right]$,
\begin{align}
&{{\mathbb{E}}_{\mathbf{W}}}\left[ \operatorname{tr}\left( {{{\mathbf{\hat{W}}}}_{k}}{\mathbf{\Sigma }_{p}}\mathbf{\hat{W}}_{k}^{\operatorname{H}}{\mathbf{\Lambda }_{i}} \right) \right] \nonumber\\
=&{{\mathbb{E}}_{\mathbf{W}}}\left[ \operatorname{tr}\left( {{{\hat{\mathbf{\Lambda }}}}_{k}}\hat{\mathbf{\Sigma }}_{p}^{-1}{\mathbf{\Sigma }_{p}}\hat{\mathbf{\Sigma }}_{p}^{-1}{{{\hat{\mathbf{\Lambda }}}}_{k}}{\mathbf{\Lambda }_{i}} \right) \right] \nonumber\\
\overset{\text{(a)}}{=}&{{\mathbb{E}}_{\mathbf{W}}}\left[ \operatorname{tr}\left( {{{\hat{\mathbf{\Lambda }}}}_{k}}\mathbf{\Sigma} _{p}^{-1/2}{{N}_\mathbf{\Sigma }}\tilde{\mathbf{\Sigma} }_{p}^{-1}{{N}_\mathbf{\Sigma }}\tilde{\mathbf{\Sigma }}_{p}^{-1}\mathbf{\Sigma} _{p}^{-1/2}{{{\hat{\mathbf{\Lambda }}}}_{k}}{\mathbf{\Lambda }_{i}} \right) \right]\nonumber \\
=&{{N}_\mathbf{\Sigma }}^{2}{{\mathbb{E}}_{\mathbf{W}}}\left[ \operatorname{tr}\left( {{{\hat{\mathbf{\Lambda }}}}_{k}}\mathbf{\Sigma} _{p}^{-1/2}\tilde{\mathbf{\Sigma }}_{p}^{-2}\mathbf{\Sigma} _{p}^{-1/2}{{{\hat{\mathbf{\Lambda}}}}_{k}}{\mathbf{\Lambda }_{i}} \right) \right] \nonumber\\
\overset{\text{(b)}}{=}&{{\mu }_{1}}{{\mathbb{E}}_\mathbf{\Lambda }}\left[ \operatorname{tr}\left( \mathbf{\Sigma} _{p}^{-1}{{{\hat{\mathbf{\Lambda} }}}_{k}}{\mathbf{\Lambda }_{i}}{{{\hat{\mathbf{\Lambda }}}}_{k}} \right) \right]\nonumber \\
\overset{\text{(c)}}{=}&{{\mu }_{1}}\operatorname{tr}\left( \mathbf{W}_{k}^{\operatorname{H}}{\mathbf{\Lambda }_{i}}{\mathbf{\Lambda }_{k}} \right)+\frac{{{\mu }_{1}}MN}{2{{N}_\mathbf{\Lambda }}}\operatorname{tr}\left( {\mathbf{\Lambda }_{i}}{\mathbf{\Sigma }_{p}} \right)\nonumber\\
&+\frac{{{\mu }_{1}}}{2{{N}_\mathbf{\Lambda }}}\operatorname{tr}\left( \mathbf{W}_{k}^{\operatorname{H}} \right)\operatorname{tr}\left( {\mathbf{\Lambda }_{i}}{\mathbf{\Lambda }_{k}} \right),
\end{align}
where (a) , (b) and (c) are obtained by substituting (\ref{s_1}) ,
(\ref{w_2}) and (\ref{a_1}), respectively.
Finally, for ${{\mathbb{E}}_{\mathbf{W}}}\left[ {{\left| \text{tr}\left( \mathbf{\hat{W}}_{k}^{\text{H}}{\mathbf{\Lambda }_{i}} \right) \right|}^{2}} \right]$,
\begin{align}
&{{\mathbb{E}}_{\mathbf{W}}}\left[ {{\left| \operatorname{tr}\left( \mathbf{\hat{W}}_{k}^{\operatorname{H}}{\mathbf{\Lambda }_{i}} \right) \right|}^{2}} \right]\nonumber\\
=&{{\mathbb{E}}_{\mathbf{W}}}\left[ {{\left| \operatorname{tr}\left( \hat{\mathbf{\Sigma }}_{p}^{-1}{{{\hat{\mathbf{\Lambda} }}}_{k}}{\mathbf{\Lambda }_{i}} \right) \right|}^{2}} \right] \nonumber\\
\overset{\text{(a)}}{=}&{{\mathbb{E}}_{\mathbf{W}}}\left[ {{\left| {{N}_\mathbf{\Sigma }}\operatorname{tr}\left( \tilde{\mathbf{\Sigma }}_{p}^{-1}\mathbf{\Sigma} _{p}^{-1/2}{{{\hat{\mathbf{\Lambda }}}}_{k}}{\mathbf{\Lambda }_{i}}\mathbf{\Sigma} _{p}^{-1/2} \right) \right|}^{2}} \right] \nonumber\\
\overset{\text{(b)}}{=}&{{\mu }_{2}}{{\mathbb{E}}_\mathbf{\Lambda }}\left[ {{\left| \operatorname{tr}\left( \mathbf{\Sigma} _{p}^{-1}{{{\hat{\mathbf{\Lambda} }}}_{k}}{\mathbf{\Lambda }_{i}} \right) \right|}^{2}} \right]\nonumber\\
&+\frac{{{\mu }_{1}}}{{{N}_\mathbf{\Sigma }}}{{\mathbb{E}}_\mathbf{\Lambda }}\left[ \operatorname{tr}\left( \mathbf{\Sigma} _{p}^{-1}{{{\hat{\mathbf{\Lambda }}}}_{k}}{\mathbf{\Lambda }_{i}}^{2}{{{\hat{\mathbf{\Lambda }}}}_{k}}\mathbf{\Sigma} _{p}^{-1} \right) \right] \nonumber\\
\overset{\text{(c)}}{=}&{{\mu }_{2}}{{\left| \operatorname{tr}\left( {\mathbf{\Lambda }_{k}}{{\mathbf{W}}_{i}} \right) \right|}^{2}}+\frac{{{\mu }_{2}}}{2{{N}_\mathbf{\Lambda }}}\operatorname{tr}\left( {{\mathbf{W}}_{i}}{\mathbf{\Sigma }_{p}}\mathbf{W}_{i}^{\operatorname{H}}{\mathbf{\Sigma }_{p}} \right)\nonumber \\
&+\frac{{{\mu }_{2}}}{2{{N}_\mathbf{\Lambda }}}\operatorname{tr}\left( {{\mathbf{W}}_{i}}{\mathbf{\Lambda }_{k}}\mathbf{W}_{i}^{\operatorname{H}}{\mathbf{\Lambda }_{k}} \right)+\frac{{{\mu }_{1}}}{{{N}_\mathbf{\Sigma }}}\operatorname{tr}\left( \mathbf{W}_{k}^{\operatorname{H}}{\mathbf{\Lambda }_{i}}^{2}{{\mathbf{W}}_{k}} \right)\nonumber \\
&+\frac{{{\mu }_{1}}MN}{2{{N}_\mathbf{\Sigma }}{{N}_\mathbf{\Lambda }}}\operatorname{tr}\left( \mathbf{\Sigma} _{p}^{-1} \right)\operatorname{tr}\left( {\mathbf{\Lambda }_{i}}^{2}{\mathbf{\Sigma }_{p}} \right)\nonumber\\
&+\frac{{{\mu }_{1}}}{2{{N}_\mathbf{\Sigma }}{{N}_\mathbf{\Lambda }}}\operatorname{tr}\left( \mathbf{\Sigma} _{p}^{-2}{\mathbf{\Lambda }_{k}} \right)\operatorname{tr}\left( {\mathbf{\Lambda }_{i}}^{2}{\mathbf{\Lambda }_{k}} \right),
\end{align}
where (a) , (b) and (c) are obtained by substituting (\ref{s_1}) ,
(\ref{w_3}) and (\ref{a_1}), respectively. 

This completes the proof.

\section{Proof of Theorem 2}\label{bp}
Similarly, using random matrix theory to solve (\ref{r_zf}) parameters.
For ${\mathbf{\Xi }^\text{th}_{p}}$, we have
\begin{align}
{{\left[ {\mathbf{\Xi }^\text{th}_{p}} \right]}_{q,j}}=&\text{tr}\left( {{{\hat{\mathbf{\Lambda} }}}_{k}}\hat{\mathbf{\Sigma }}_{p}^{-1}\hat{\mathbf{\Sigma}}_{p}^{-1}{{{\hat{\mathbf{\Lambda}}}}_{i}}{\mathbf{\Sigma }_{p}} \right) \nonumber\\
\overset{\text{(a)}}{=}&\text{tr}\left( {{{\hat{\mathbf{\Lambda}}}}_{k}}\mathbf{\Sigma} _{p}^{-1/2}{{N}_\mathbf{\Sigma }}\tilde{\mathbf{\Sigma}}_{p}^{-1}{{N}_\mathbf{\Sigma }}\tilde{\mathbf{\Sigma}}_{p}^{-1}\mathbf{\Sigma} _{p}^{-1/2}{{{\hat{\mathbf{\Lambda}}}}_{i}}{\mathbf{\Sigma }_{p}} \right)\nonumber \\
=&N_\mathbf{\Sigma }^{2}\text{tr}\left( {{{\hat{\mathbf{\Lambda}}}}_{k}}\mathbf{\Sigma} _{p}^{-1/2}\tilde{\mathbf{\Sigma}}_{p}^{-2}\mathbf{\Sigma} _{p}^{-1/2}{{{\hat{\mathbf{\Lambda}}}}_{i}} \right) \nonumber\\
\overset{\text{(b)}}{=}&{{\mu }_{1}}\text{tr}\left( {{{\hat{\mathbf{\Lambda}}}}_{k}}\mathbf{\Sigma} _{p}^{-1}{{{\hat{\mathbf{\Lambda}}}}_{i}} \right) \nonumber\\
\overset{\text{(c)}}{=}&{{\mu }_{1}}\text{tr}\left( {\mathbf{\Lambda }_{k}}\mathbf{\Sigma} _{p}^{-1}{\mathbf{\Lambda }_{i}} \right)+\frac{{{\mu }_{1}}MN}{2{{N}_\mathbf{\Lambda }}}\text{tr}\left( {\mathbf{\Sigma }_{p}} \right)\nonumber\\
&+\frac{{{\mu }_{1}}}{2{{N}_\mathbf{\Lambda }}}\text{tr}\left( {\mathbf{\Lambda }_{k}} \right)\text{tr}\left( \mathbf{\Sigma} _{p}^{-1}{\mathbf{\Lambda }_{i}} \right),
\end{align}
where (a) , (b) and (c) result from (\ref{s_1}) ,
(\ref{w_2}) and (\ref{a_1}), respectively. Follow the same steps as above, we have
\begin{align}
{{\left[ {{{\tilde{\mathbf{\Xi }}}}^\text{th}_{p}} \right]}_{q,j}}=&\text{tr}\left( {{{\hat{\mathbf{\Lambda}}}}_{k}}\hat{\mathbf{\Sigma}}_{p}^{-1}\hat{\mathbf{\Sigma}}_{p}^{-1}{{{\hat{\mathbf{\Lambda}}}}_{i}}{\mathbf{\Sigma }_{p}}\tilde{\mathbf{\Gamma}} \right) \nonumber\\
\overset{\text{(a)}}{=}&\text{tr}\left( {{{\hat{\mathbf{\Lambda }}}}_{k}}\mathbf{\Sigma} _{p}^{-1/2}{{N}_\mathbf{\Sigma }}\tilde{\mathbf{\Sigma}}_{p}^{-1}{{N}_\mathbf{\Sigma }}\tilde{\mathbf{\Sigma} }_{p}^{-1}\mathbf{\Sigma} _{p}^{-1/2}{{{\hat{\mathbf{\Lambda}}}}_{i}}{\mathbf{\Sigma }_{p}}\tilde{\mathbf{\Gamma}} \right) \nonumber\\
=&N_\mathbf{\Sigma}^{2}\text{tr}\left({{{\hat{\mathbf{\Lambda}}}}_{k}}\mathbf{\Sigma}_{p}^{-1/2}\tilde{\mathbf{\Sigma}}_{p}^{-2}\mathbf{\Sigma}_{p}^{-1/2}{{{\hat{\mathbf{\Lambda}}}}_{i}}\tilde{\mathbf{\Gamma}} \right) \nonumber\\
\overset{\text{(b)}}{=}&{{\mu}_{1}}\text{tr}\left({{{\hat{\mathbf{\Lambda}}}}_{k}}\mathbf{\Sigma} _{p}^{-1}{{{\hat{\mathbf{\Lambda }}}}_{i}}\tilde{\mathbf{\Gamma }}\right)\nonumber\\
\overset{\text{(c)}}{=}&{{\mu }_{1}}\text{tr}\left( {\mathbf{\Lambda }_{k}}\mathbf{\Sigma} _{p}^{-1}{\mathbf{\Lambda }_{i}}\tilde{\mathbf{\Gamma}} \right)+\frac{{{\mu }_{1}}MN}{2{{N}_\mathbf{\Lambda }}}\text{tr}\left( {\mathbf{\Sigma }_{p}}\tilde{\mathbf{\Gamma}} \right) \nonumber\\
&+\frac{{{\mu }_{1}}}{2{{N}_\mathbf{\Lambda }}}\text{tr}\left( {\mathbf{\Lambda }_{k}}\tilde{\mathbf{\Gamma}} \right)\text{tr}\left( \mathbf{\Sigma} _{p}^{-1}{\mathbf{\Lambda }_{i}} \right), 
\end{align}
where (a) is obtained by substituting (\ref{s_1}), (b) results from (\ref{w_2}) and (c) results from (\ref{a_1}). Here, we regard the interference $\tilde{\mathbf{\Gamma}}$ as an independent variable, and we also can get approximately accurate results. So the variable $\tilde{\mathbf{\Gamma}}$ is
\begin{equation}
\tilde{\mathbf{\Gamma}}=\sum\limits_{k=1}^{K}{\left[ {\mathbf{\Lambda }_{k}}-\mathcal{E}^{cov}\right]}+{\mathbf{\sigma }^{2}}{{\mathbf{I}}_{MN}},
\end{equation}
where
\begin{align}
\mathcal{E}^{cov}=&{{{\hat{\mathbf{\Lambda }}}}_{k}}\hat{\mathbf{\Sigma}}_{p}^{-1}\hat{\mathbf{\Sigma}}_{p}^{-1}{{{\hat{\mathbf{\Lambda}}}}_{k}}{\mathbf{\Sigma }_{p}} \nonumber\\
\overset{\text{(a)}}{=}&{\mathbf{\Lambda }_{k}}\hat{\mathbf{\Sigma }}_{p}^{-1}\hat{\mathbf{\Sigma }}_{p}^{-1}{\mathbf{\Lambda }_{k}}{\mathbf{\Sigma }_{p}}+\frac{1}{2{{N}_\mathbf{\Lambda }}}\mathbf{\Sigma} _{p}^{2}\text{tr}\left( \hat{\mathbf{\Sigma}}_{p}^{-1}\hat{\mathbf{\Sigma}}_{p}^{-1}{\mathbf{\Sigma }_{p}} \right)\nonumber\\
&+\frac{1}{2{{N}_\mathbf{\Lambda }}}{\mathbf{\Lambda }_{k}}{\mathbf{\Sigma }_{p}}\text{tr}\left( \hat{\mathbf{\Sigma}}_{p}^{-1}\hat{\mathbf{\Sigma}}_{p}^{-1}{\mathbf{\Lambda }_{k}} \right)\nonumber\\
\overset{\text{(b)}}{=}&N_{\mathbf{\Sigma}}^{2}{\mathbf{\Lambda }_{k}}\mathbf{\Sigma} _{p}^{-1/2}\tilde{\mathbf{\Sigma} }_{p}^{-1}\mathbf{\Sigma} _{p}^{-1}\tilde{\mathbf{\Sigma}}_{p}^{-1}\mathbf{\Sigma} _{p}^{-1/2}{\mathbf{\Lambda }_{k}}{\mathbf{\Sigma }_{p}}\nonumber\\
&+\frac{{{\mu }_{1}}}{2{{N}_\mathbf{\Lambda }}}\mathbf{\Sigma} _{p}^{2}\text{tr}\left( \mathbf{\Sigma} _{p}^{-1} \right)+\frac{{{\mu }_{1}}}{2{{N}_\mathbf{\Lambda }}}{\mathbf{\Lambda }_{k}}{\mathbf{\Sigma }_{p}}\text{tr}\left( {\mathbf{\Lambda }_{k}} \right),
\end{align}
(a) results from (\ref{a_1}), and (b) from (\ref{s_1}). 

This completes the proof.

\ifCLASSOPTIONcaptionsoff
\newpage
\fi

\bibliographystyle{IEEEtran}
\bibliography{references}

\begin{thebibliography}{10}
\providecommand{\url}[1]{#1}
\csname url@samestyle\endcsname
\providecommand{\newblock}{\relax}
\providecommand{\bibinfo}[2]{#2}
\providecommand{\BIBentrySTDinterwordspacing}{\spaceskip=0pt\relax}
\providecommand{\BIBentryALTinterwordstretchfactor}{4}
\providecommand{\BIBentryALTinterwordspacing}{\spaceskip=\fontdimen2\font plus
\BIBentryALTinterwordstretchfactor\fontdimen3\font minus
  \fontdimen4\font\relax}
\providecommand{\BIBforeignlanguage}[2]{{%
\expandafter\ifx\csname l@#1\endcsname\relax
\typeout{** WARNING: IEEEtran.bst: No hyphenation pattern has been}%
\typeout{** loaded for the language `#1'. Using the pattern for}%
\typeout{** the default language instead.}%
\else
\language=\csname l@#1\endcsname
\fi
#2}}
\providecommand{\BIBdecl}{\relax}
\BIBdecl

\bibitem{shafi20175g}
M.~Shafi, A.~F. Molisch, P.~J. Smith, T.~Haustein, P.~Zhu, P.~De~Silva,
  F.~Tufvesson, A.~Benjebbour, and G.~Wunder, ``{5G: A tutorial overview of
  standards, trials, challenges, deployment, and practice},'' \emph{IEEE
  journal on selected areas in communications}, vol.~35, no.~6, pp. 1201--1221,
  2017.

\bibitem{ngo2017cell}
H.~Q. Ngo, A.~Ashikhmin, H.~Yang, E.~G. Larsson, and T.~L. Marzetta,
  ``{Cell-free massive MIMO versus small cells},'' \emph{IEEE Transactions on
  Wireless Communications}, vol.~16, no.~3, pp. 1834--1850, 2017.

\bibitem{interdonato2019ubiquitous}
G.~Interdonato, E.~Bj{\"o}rnson, H.~Q. Ngo, P.~Frenger, and E.~G. Larsson,
  ``{Ubiquitous cell-free massive MIMO communications},'' \emph{EURASIP Journal
  on Wireless Communications and Networking}, vol. 2019, no.~1, pp. 1--13,
  2019.

\bibitem{chen2017can}
Z.~Chen and E.~Bj{\"o}rnson, ``{Can we rely on channel hardening in cell-free
  massive MIMO?}'' in \emph{2017 IEEE Globecom Workshops (GC Wkshps)}.\hskip
  1em plus 0.5em minus 0.4em\relax IEEE, 2017, pp. 1--6.

\bibitem{buzzi2017cell}
S.~Buzzi and C.~D’Andrea, ``{Cell-free massive MIMO: User-centric
  approach},'' \emph{IEEE Wireless Communications Letters}, vol.~6, no.~6, pp.
  706--709, 2017.

\bibitem{ngo2017total}
H.~Q. Ngo, L.-N. Tran, T.~Q. Duong, M.~Matthaiou, and E.~G. Larsson, ``{On the
  total energy efficiency of cell-free massive MIMO},'' \emph{IEEE Transactions
  on Green Communications and Networking}, vol.~2, no.~1, pp. 25--39, 2017.

\bibitem{zhang2017spectral}
J.~Zhang, Y.~Wei, E.~Bj{\"o}rnson, Y.~Han, and X.~Li, ``{Spectral and energy
  efficiency of cell-free massive MIMO systems with hardware impairments},'' in
  \emph{2017 9th International Conference on Wireless Communications and Signal
  Processing (WCSP)}.\hskip 1em plus 0.5em minus 0.4em\relax IEEE, 2017, pp.
  1--6.

\bibitem{bashar2018cell}
M.~Bashar, K.~Cumanan, A.~G. Burr, H.~Q. Ngo, and M.~Debbah, ``{Cell-free
  massive MIMO with limited backhaul},'' in \emph{2018 IEEE International
  Conference on Communications (ICC)}.\hskip 1em plus 0.5em minus 0.4em\relax
  IEEE, 2018, pp. 1--7.

\bibitem{nguyen2017energy}
L.~D. Nguyen, T.~Q. Duong, H.~Q. Ngo, and K.~Tourki, ``{Energy efficiency in
  cell-free massive MIMO with zero-forcing precoding design},'' \emph{IEEE
  Communications Letters}, vol.~21, no.~8, pp. 1871--1874, 2017.

\bibitem{hoang2018cell}
T.~M. Hoang, H.~Q. Ngo, T.~Q. Duong, H.~D. Tuan, and A.~Marshall, ``{Cell-free
  massive MIMO networks: Optimal power control against active eavesdropping},''
  \emph{IEEE Transactions on Communications}, vol.~66, no.~10, pp. 4724--4737,
  2018.

\bibitem{mai2018cell}
T.~C. Mai, H.~Q. Ngo, and T.~Q. Duong, ``{Cell-free massive MIMO systems with
  multi-antenna users},'' in \emph{2018 IEEE Global Conference on Signal and
  Information Processing (GlobalSIP)}.\hskip 1em plus 0.5em minus 0.4em\relax
  IEEE, 2018, pp. 828--832.

\bibitem{dao2020effective}
H.~T. Dao and S.~Kim, ``{Effective channel gain-based access point selection in
  cell-free massive MIMO systems},'' \emph{IEEE Access}, vol.~8, pp.
  108\,127--108\,132, 2020.

\bibitem{jin2019channel}
Y.~Jin, J.~Zhang, S.~Jin, and B.~Ai, ``{Channel estimation for cell-free mmWave
  massive MIMO through deep learning},'' \emph{IEEE Transactions on Vehicular
  Technology}, vol.~68, no.~10, pp. 10\,325--10\,329, 2019.

\bibitem{wang2020uplink}
Z.~Wang, J.~Zhang, E.~Bj{\"o}rnson, and B.~Ai, ``{Uplink Performance of
  Cell-Free Massive MIMO Over Spatially Correlated Rician Fading Channels},''
  \emph{IEEE Communications Letters}, 2020.

\bibitem{mishra2015analysis}
S.~Mishra, R.~Prusty, and P.~K. Hota, ``{Analysis of Levenberg-Marquardt and
  Scaled Conjugate gradient training algorithms for artificial neural network
  based LS and MMSE estimated channel equalizers},'' in \emph{2015
  International Conference on Man and Machine Interfacing (MAMI)}.\hskip 1em
  plus 0.5em minus 0.4em\relax IEEE, 2015, pp. 1--7.

\bibitem{bjornson2016massive}
E.~Bj{\"o}rnson, L.~Sanguinetti, and M.~Debbah, ``{Massive MIMO with imperfect
  channel covariance information},'' in \emph{2016 50th Asilomar Conference on
  Signals, Systems and Computers}.\hskip 1em plus 0.5em minus 0.4em\relax IEEE,
  2016, pp. 974--978.

\bibitem{yin2013coordinated}
H.~Yin, D.~Gesbert, M.~Filippou, and Y.~Liu, ``{A coordinated approach to
  channel estimation in large-scale multiple-antenna systems},'' \emph{IEEE
  Journal on selected areas in communications}, vol.~31, no.~2, pp. 264--273,
  2013.

\bibitem{upadhya2018covariance}
K.~Upadhya and S.~A. Vorobyov, ``{Covariance matrix estimation for massive
  MIMO},'' \emph{IEEE Signal Processing Letters}, vol.~25, no.~4, pp. 546--550,
  2018.

\bibitem{neumann2018covariance}
D.~Neumann, M.~Joham, and W.~Utschick, ``{Covariance matrix estimation in
  massive MIMO},'' \emph{IEEE Signal Processing Letters}, vol.~25, no.~6, pp.
  863--867, 2018.

\bibitem{neumann2018learning}
D.~Neumann, T.~Wiese, and W.~Utschick, ``{Learning the MMSE channel
  estimator},'' \emph{IEEE Transactions on Signal Processing}, vol.~66, no.~11,
  pp. 2905--2917, 2018.

\bibitem{interdonato2020self}
G.~Interdonato, P.~Frenger, and E.~G. Larsson, ``{Self-Learning detector for
  the cell-Free massive MIMO uplink: The line-of-sight case},'' in \emph{2020
  IEEE 21st International Workshop on Signal Processing Advances in Wireless
  Communications (SPAWC)}.\hskip 1em plus 0.5em minus 0.4em\relax IEEE, 2020,
  pp. 1--5.

\bibitem{haghighatshoar2017massive}
S.~Haghighatshoar and G.~Caire, ``{Massive MIMO pilot decontamination and
  channel interpolation via wideband sparse channel estimation},'' \emph{IEEE
  Transactions on Wireless Communications}, vol.~16, no.~12, pp. 8316--8332,
  2017.

\bibitem{marzetta2010noncooperative}
T.~L. Marzetta, ``{Noncooperative cellular wireless with unlimited numbers of
  base station antennas},'' \emph{IEEE transactions on wireless
  communications}, vol.~9, no.~11, pp. 3590--3600, 2010.

\bibitem{jose2011pilot}
J.~Jose, A.~Ashikhmin, T.~L. Marzetta, and S.~Vishwanath, ``{Pilot
  contamination and precoding in multi-cell TDD systems},'' \emph{IEEE
  Transactions on Wireless Communications}, vol.~10, no.~8, pp. 2640--2651,
  2011.

\bibitem{pitarokoilis2017effect}
A.~Pitarokoilis, E.~Bj{\"o}mson, and E.~G. Larsson, ``{On the effect of
  imperfect timing synchronization on pilot contamination},'' in \emph{2017
  IEEE International Conference on Communications (ICC)}.\hskip 1em plus 0.5em
  minus 0.4em\relax IEEE, 2017, pp. 1--6.

\bibitem{cao2018uplink}
J.~Cao, D.~Wang, J.~Li, Q.~Sun, and Y.~Hu, ``Uplink spectral efficiency
  analysis of multi-cell multi-user massive {MIMO} over correlated ricean
  channel,'' \emph{Science China Information Sciences}, vol.~61, no.~8, p.
  082305, 2018.

\bibitem{marzetta2016fundamentals}
T.~L. Marzetta and H.~Q. Ngo, \emph{Fundamentals of massive MIMO}.\hskip 1em
  plus 0.5em minus 0.4em\relax Cambridge University Press, 2016.

\bibitem{tulino2004random}
A.~M. Tulino, S.~Verd{\'u}, and S.~Verdu, \emph{Random matrix theory and
  wireless communications}.\hskip 1em plus 0.5em minus 0.4em\relax Now
  Publishers Inc, 2004.

\end{thebibliography}
\end{document}